\documentclass[12pt]{article}

\pdfoutput=1
\usepackage{jheppub}

\newcommand{\CN}{{\cal N}}
\newcommand{\CO}{{\cal O}}
\newcommand{\CB}{{\cal B}}
\newcommand{\CM}{{\cal M}}
\newcommand{\CZ}{{\cal Z}}
\newcommand{\CX}{{\cal X}}
\newcommand{\CF}{{\cal F}}
\newcommand{\CL}{{\cal L}}
\newcommand{\CV}{{\cal V}}
\newcommand{\CI}{{\cal I}}

\newcommand{\CD}{{\cal D}}

\newcommand{\CA}{{\cal A}}

\newcommand{\C}{{\mathbb C}}

\DeclareMathOperator{\Bun}{Bun}
\DeclareMathOperator{\Loc}{Loc}
\DeclareMathOperator{\Hom}{Hom}
\DeclareMathOperator{\End}{End}
\DeclareMathOperator{\Jac}{Jac}
\DeclareMathOperator{\Pic}{Pic}
\newcommand{\fD}{{\mathfrak D}}
\newcommand{\fB}{{\mathfrak B}}

\title{S-duality of boundary conditions and the Geometric Langlands program}
\abstract{Maximally supersymmetric gauge theory in four dimensions admits local boundary conditions which preserve half of the bulk supersymmetries. 
The S-duality of the bulk gauge theory can be extended in a natural fashion to act on such half-BPS boundary conditions.
The purpose of this note is to explain the role these boundary conditions can play in the Geometric Langlands program. 
In particular, we describe how to obtain pairs of 
Geometric Langland dual objects from S-dual pairs of half-BPS boundary conditions.}\author[1]{Davide Gaiotto}
\affiliation[1]{Perimeter Institute for Theoretical Physics, Waterloo, Ontario, Canada N2L 2Y5}

\begin{document}
\maketitle
\flushbottom
\section{Introduction and conclusions}
The beautiful work of \cite{Kapustin:2006pk} established a rich dictionary between S-duality in four-dimensional $\CN=4$ gauge theory 
and the mathematical subject of Geometric Langlands duality. We refer to that paper and to the more recent work \cite{Witten:2015aa} 
for an extensive discussion of the relationship and references to the mathematical literature on the subject. 

The Geometric Langlands duality is a relation between two categories associated to a Riemann surface $C$ and Langlands dual Lie groups 
$G$ and ${}^\vee G$. These two categories are interpreted as an abstract description of two categories of boundary conditions (``branes'') 
for a two-dimensional physical system, the twisted compactification of $\CN=4$ gauge theory on $C$. In turn, the duality between the two categories is a form of two-dimensional mirror symmetry 
which follows from the S-duality between the $G$ and ${}^\vee G$ gauge theories \cite{Bershadsky:1995vm}. 

A natural way to define boundary conditions in two-dimensions is to start from a local boundary condition 
in four-dimensional $\CN=4$ gauge theory, compactified along $C$. This was already crucial in the original work on the subject \cite{Kapustin:2006pk}:
half-BPS Dirichlet boundary conditions in the four-dimensional gauge theory map to
the most basic objects on one side of the GL duality, labelled by a $G$ local system on $C$. 
The S-duality image ${}^\vee \fD$ of Dirichlet boundary conditions
should then map to the GL dual image of these objects, the so-called ``Hecke eigensheaves''.

This relation is explained in \cite{Kapustin:2006pk} with the help of half-BPS line defects in the four-dimensional gauge theory. 
Half-BPS line defects inserted at points on $C$ map to interfaces in the two dimensional system and thus to interesting functors acting on the 
categories of branes. Wilson lines act on the Chan-Paton bundle of branes in a simple manner, by tensor product with a certain reference bundle.
'T Hooft operators act by Hecke correspondences, functors which play a key role in Geometric Langlands. 

Dirichlet boundary conditions $\fD$ produce ``eigenbranes'' for the action of Wilson lines, simply because they fix the value at the boundary of the 
connection and scalar fields which enter the Wilson line path-ordered exponential. S-duality exchanges Wilson lines and 't Hooft lines. 
As a consequence, the S-dual boundary condition ${}^\vee \fD$ will produce ``eigenbranes'' for the action of  't Hooft lines, i.e. Hecke eigensheaves.

The analysis of \cite{Kapustin:2006pk} did not employ directly the S-dual boundary condition ${}^\vee \fD$, as
the action of S-duality on boundary conditions for non-Abelian gauge theory was not understood at the time. 
This provided an important motivation for the work of \cite{Gaiotto:2008sa,Gaiotto:2008ak}, 
which described general half-BPS boundary conditions four-dimensional $\CN=4$ gauge theory and their behaviour under the action of S-duality.

One conclusion of that work was that ${}^\vee \fD$ can be defined with the help of a very special three-dimensional $\CN=4$ SCFT $T[G]$, 
which unfortunately lacks a simple UV Lagrangian description. For some classical groups a UV Lagrangian description is available, but breaks 
symmetries which are crucial for GL applications. On the other hand, the analysis also provided us with a large list of S-dual pairs of boundary conditions 
and interfaces which do admit an effective UV Lagrangian description. 

The purpose of this paper is to make explicit the map from half-BPS boundary conditions which admit an effective Lagrangian description to 
objects in the categories relevant for the Geometric Langlands duality (see \cite{Henningson:2011qk} for some early work on the subject). 
The expected payoff is a long list of Geometric Langlands dual pairs of mathematical objects which can be defined uniformly on any Riemann surface. 

The obvious limitation of the computational techniques employed in this paper is the requirement of an effective UV description of the boundary conditions. 
This prevents us from, say, constructing generic Hecke eigensheaves directly from ${}^\vee \fD$. 
In a separate paper, \cite{Next}, we will discuss alternative computational approaches based on Vertex Operator Algebras, 
which appear to have a broader range of applicability.

\subsection{The twisted compactification of half-BPS boundary conditions}
We employ a standard twisted compactification of maximally supersymmetric four-dimensional gauge theory with some gauge group $G$ on a Riemann surface $C$. 
The R-symmetry of the theory is $SO(6)_R$, which has an obvious $SO(2)_V \times SO(4)_A$ block-diagonal subgroup. 
The twist on $C$ involves the $SO(2)_V$ factor in the subgroup. The $SO(4)_A$ factor survives the twisting and becomes 
an R-symmetry group of the compactified theory. 
 
The twisted compactification preserves $(4,4)$ two-dimensional supersymmetry. 
After the twist, the fields of the four-dimensional gauge theory can be 
organized into a two-dimensional $(4,4)$ gauged linear sigma model 
with infinite-dimensional target space and gauge group: the target space is the cotangent bundle 
to the space of $G$ connections on $C$ and the gauge group consists of maps from $C$ to $G$. 

Concretely, we denote the connection on $C$ as $A$ and the $(1,0)$ and $(0,1)$ components of 
the corresponding covariant derivative as $\partial_A$ and $\bar \partial_A$. We denote the $(1,0)$ and $(0,1)$ components of 
the Higgs field (the scalar superpartner parameterizing the cotangent directions) as $\Phi$, $\bar \Phi$. 

\subsubsection{Hitchin equations}
An important branch of the moduli space of 2d supersymmetric vacua of the system is given by the space of 
solutions $\CM_H(C,G)$ of $G$ Hitchin equations on $C$ modulo gauge transformations. 
Recall that Hitchin equations are two-dimensional differential equations for a connection and Higgs field on $C$ \cite{Hitchin:1986vp}: 
\begin{equation}
\bar \partial_A \Phi = 0 \qquad \qquad F_A + [\Phi, \bar \Phi]=0
\end{equation}
The moduli space $\CM_H(C,G)$ is an hyper-K\"ahler manifold. 

For most purposes, the result of the twisted compactification can be described as a two-dimensional non-linear sigma model with target $\CM_H(C,G)$.
The underlying gauge theory description may become important at special loci in $\CM_H(C,G)$ where some gauge symmetry is unbroken \cite{Witten:2008ep}. 
Whenever possible, we will focus on situations for which the $\CM_H(C,G)$
sigma model description is adequate.

Recall that a hyper-K\"ahler manifold can be endowed with a $\mathbb{P}^1$ worth of different complex structures, 
the ``twistor sphere'', parameterized by a complex variable $\zeta$. 

In complex structure ``I'', i.e. $\zeta = 0$, $\CM_H(C,G)$ can be identified with the moduli space of Higgs bundles, i.e. 
pairs $(\bar \partial_A, \Phi)$ with 
\begin{equation}
\bar \partial_A \Phi =0
\end{equation}
modulo complexified gauge transformations. A conjugate description holds at the opposite complex structure $\bar I$, i.e. $\zeta = \infty$.

In all other complex structures, $\CM_H(C,G)$ can be identified with a space $\Loc(C,G)$
of complexified $G$ flat connections modulo complexified gauge transformations, 
sending the pair $(A,\Phi)$ into the Lax connection 
\begin{equation}
D^\zeta = \partial_A + \zeta^{-1} \Phi \qquad \qquad \bar D^\zeta = \bar \partial_A + \zeta \bar \Phi
\end{equation} 

The S-duality relation between $G$ and ${}^\vee G$ four-dimensional gauge theories maps 
to a mirror symmetry relation between $\CM_H(C,G)$ and $\CM_H(C,{}^\vee G)$. This is a mirror symmetry 
in the sense of SYZ \cite{Strominger:1996it}: both spaces are complex integrable systems in complex structure $I$ with the same base, parameterized by 
gauge-invariant polynomials built from the Higgs fields $\Phi$ and ${}^\vee \Phi$. 
The non-singular fibers are dual Abelian varieties. 

Mirror symmetry relates BPS boundary conditions in the two sigma models. The Geometric Langlands program is expected to 
encode mathematical aspects of the mirror relationship between two specific classes of half-BPS boundary conditions: BBB and BAA branes. 
\footnote{Mirror symmetry and the Geometric Langlands program can also be thought of as matching the larger categories
of B and A branes in a specific complex structure ``$K$''. The sub-categories of BBB and BAA branes, though, have useful extra structures 
and occur most naturally in our setup.}

\subsubsection{BBB branes}
BBB branes are boundary conditions which are compatible with the B-model twist in 
every complex structure of $\CM_H(C,G)$. A prototypical BBB brane is defined by a Chan-Paton bundle on 
$\CM_H(C,G)$ equipped with a hyper-holomorphic connection, i.e. a connection whose curvature is of type $(1,1)$ 
in all complex structures. 

More general BBB branes should be described by some hyper-holomorphic version of 
complexes of coherent sheaves, as a BBB brane will map to a standard B-brane in every complex structure. 
We do not know a necessary set of mathematical conditions for such a family of 
B-branes over the twistor sphere to give a BBB brane, but in Appendices \ref{app:SQM} and \ref{app:category} we will discuss 
important examples of BBB branes and some tentative characterization of a general story. 

The ``classical'' discussion of the mirror symmetry transformation of BBB branes is easiest in complex structure $I$, 
but in standard Geometric Langlands applications the description in other complex structures becomes important. 

The basic BBB objects in Geometric Langlands are Dirichlet boundary conditions $D_\mu$, which constrain the fields at the boundary to lie at a specific point $\mu$ in the 
Hitchin moduli space. In particular, they are labelled by a point in $\Loc(C,G)$.

\subsubsection{BAA branes}
BAA branes are boundary conditions which are compatible with B twist in complex structure $I$ 
and A-twist in complex structures orthogonal to $I$, such as $J$ and $K$. A prototypical BAA brane
is supported on a complex Lagrangian submanifold in the moduli space of Higgs bundles, 
equipped with a Chan-Paton bundle with a connection which is both flat and holomorphic in complex structure $I$.

In standard Geometric Langlands applications, one roughly identifies $\CM_H(C,G)$ with the cotangent bundle to 
the space $\Bun(C,G)$ of $G$ bundles on $C$ and ``quantizes'' the BAA brane to a (twisted) D-module 
on $\Bun(C,G)$. The quantization is explained in \cite{Kapustin:2006pk} through the notion of a ``canonical coisotropic brane'' 
(see also \cite{Gukov:2008ve,Gukov:2010sw}).

Another useful perspective based on $\CN=4$ supersymmetric quantum mechanics (see Appendix \ref{app:SQM}) can be given 
with the help of auxiliary BAA branes $F_p$ supported on the fiber of the cotangent bundle over a point $p$ of $\Bun(C,G)$:
the stalks of the sheaf underlying the D-module consist of the spaces $\Hom_A(F_p,B)$ of A-model morphisms from $F_p$ to the BAA brane $B$ in consideration. 
\footnote{Yet another perspective on this relation (again discussed in Appendix \ref{app:SQM}) is that general BAA branes in a 
neighbourhood of $\Bun(C,G)$ in $\CM_H(C,G)$ can be parameterized by a solution $(\CA, \Phi)$ of generalized Hitchin equations 
on $\Bun(C,G)$. The connection $\CA$ is interpreted as the connection of a Chan-Paton bundle for a brane wrapping $\Bun(C,G)$
and $\Phi$ as a matrix-valued transverse deformation of the shape of the BAA brane. In complex structure $I$ 
the spectral data of the pair $(\CA, \Phi)$ gives a complex Lagrangian submanifold in $T^*\Bun(C,G)$
equipped with a line bundle. In general complex structure, the flat Lax connection $D_\CA + \zeta^{-1} \Phi$ and 
$\bar D_\CA + \zeta \bar \Phi$ can be interpreted as a holomorphic D-module on $\Bun(C,G)$.}

\subsubsection{Twisted compactification}

A natural way to produce BBB and BAA branes in the two-dimensional theory is to adapt the twisted compactification along $C$ to standard half-BPS boundary conditions
for the four-dimensional gauge theory. \footnote{Another natural way is to consider surface defects in class S theories \cite{Gaiotto:2011tf}.}
These boundary conditions naturally preserve an $SO(3)_H \times SO(3)_C$ subgroup 
of $SO(6)_R$. Embedding $SO(2)_V$ as a Cartan subgroup in $SO(3)_H$ or $SO(3)_C$ respectively, we obtain 
either a BAA or a BBB brane in the two-dimensional theory. 

For example, starting from four-dimensional Dirichlet boundary conditions $\fD$, the BAA twist produces the $F_p$ branes 
mentioned above, while the BBB twist produces Dirichlet boundary conditions $D_\mu$.

S-duality acts on boundary conditions in a non-trivial manner, while also exchanging the two $SO(3)$ R-symmetry factors. 
Each pair $(\fB, {}^\vee \fB)$ of S-dual four-dimensional boundary conditions will map to two pairs of mirror 2d boundary conditions: the 
BAA twist of $\fB$ is dual to the BBB twist of ${}^\vee \fB$ and vice-versa. 

In particular, the S-dual of Dirichlet boundary conditions ${}^\vee \fD$ would map under BAA twist to the Hecke eigensheaves 
${}^\vee D_\mu$ dual to the BBB Dirichlet boundary conditions, while it would map under the BBB twist to some BBB branes ${}^\vee F_p$. 
The latter branes should have the property that B-model morphisms $\Hom_B({}^\vee F_p,E)$
are the stalks of the D-module dual to any sheaf $E$ on $\Loc(C,G)$, i.e. should be ``kernels'' for the Geometric Langlands duality.

In most of the paper we will discuss the ``classical'' descriptions of the BAA and BBB branes associated to $\fB$ 
as complex Lagrangian manifolds $\CL(C,G,\fB)$ and sheaves $\CV(C,G,\fB)$ in the moduli space of Higgs bundles. 
In particular, we will test the expectation that $\CL(C,G,\fB)$ and $\CV(C,{}^\vee G,{}^\vee \fB)$ should 
be related by fiberwise mirror-symmetry at generic points in the base of the complex integrable system. 

At the end of the paper we will discuss more briefly the corresponding D-modules on $\Bun(C,G)$ and the sheaves 
on $\Loc(C,G)$, leaving further discussion to the companion paper \cite{Next}.

\subsubsection{Symmetries and background fields}
In some situations, a single half-BPS boundary condition in four dimensions will map to a {\it family} 
of BAA and BBB boundary conditions in two dimensions. This will occur if the four-dimensional boundary condition 
is equipped with extra global symmetries, which can be coupled to background gauge multiplets on $C$
in the twisted compactification process. 

On general grounds, an half-BPS boundary condition may admit two types of global symmetry groups, exchanged by 
S-duality. The group of ``Higgs branch'' symmetries $G_H$ can be coupled to background vectormultiplets, 
meaning a $G_H$ bundle for a BAA twist or a $G_H$ Hitchin system for a BBB twist. 
The group of ``Coulomb branch'' symmetries $G_C$ can be coupled to background twisted vectormultiplets, 
meaning a $G_C$ bundle for a BBB twist or a $G_C$ Hitchin system for a BAA twist.

As a result, a single four-dimensional boundary condition will give us a family of 
two-dimensional BAA boundary conditions parameterized by $\Bun(G_H,C) \times \CM_H(G_C,C)$
or a family of two-dimensional BBB boundary conditions parameterized by $\Bun(G_C,C) \times \CM_H(G_H,C)$.

S-duality will then relate the two families associated to a pair of S-dual four-dimensional boundary conditions. 

\subsection{General strategy}
Maximally supersymmetric four-dimensional gauge theory admits a large variety of half-BPS boundary conditions. We adopt a divide-and-conquer 
strategy, focussing our analysis on simple boundary conditions and interfaces, which can be concatenated to produce more general examples. 
In particular, all boundary conditions which admit an explicit Lagrangian description can be decomposed in such a manner. 
Concatenation of interfaces in the four-dimensional gauge theory will map to concatenation of interfaces in the two-dimensional setup and then to 
appropriate convolution operations in the categories of branes. 

The most important and novel part of our analysis concerns boundary conditions and interfaces where the 
four-dimensional gauge fields are coupled to three-dimensional hypermultiplets in some symplectic representation $M$ of $G$. 

The BAA twist of such boundary conditions will give us a class of complex Lagrangian submanifolds $\CL(C,G,M)$ of the space 
of Higgs bundles labelled by $M$. The mathematical treatment of such Lagrangians has been initiated recently by 
\cite{Hitchin:2016aa}. These Lagrangian submanifolds admit a natural quantization 
to D-modules given by conformal blocks for a free symplectic boson vertex algebra. 

The BBB twist of such boundary conditions gives us a hyper-holomorphic bundle $\CV(C,G,M)$ built as the ``quantization'' of 
the Dirac-Higgs bundle, itself defined as an hyper-holomorphic bundle of 
zeromodes of a Dirac operator couples to a solution of the Hitchin system (See \cite{dirachiggs} for a review of the Dirac-Higgs bundle). 

The corresponding sheaf in any complex structure can be computed rather explicitly as the cohomology of a complex 
of forms on $C$ valued in the bundle associated to $M$.

Other useful ingredients are various generalization of Dirichlet boundary conditions, which we review in Appendix \ref{sec:dir}.

\subsection{Open problems and future directions}
Even in the simplest examples discussed in this paper, our analysis 
barely scratches the surface of very rich mathematical structures, 
which should be explored in further detail. Many interesting calculations are left for future work. 

Line defects play a prominent role in the analysis of Kapustin and Witten and in the comparison with Geometric Langlands. 
Boundary conditions on four-dimensional gauge theory may support Wilson-like and vortex-like 
BPS line defects related by S-duality, which can be added at points in $C$ during the twisted compactification
in order to modify the resulting  BAA and BBB branes. The S-duality map should be a generalization of the mirror map discussed in 
detail in \cite{Assel:2015oxa}.

Bulk line defects brought to the boundary will map to boundary line defects. This may allow one to systematically discuss
the effect of Hecke modifications on the D-modules defined in this paper. Boundary line defects will also arise as the endpoint of bulk surface defects, which occur in the study of 
Geometric-Langlands duality for ramified Hitchin systems \cite{Gukov:2006jk}.
Work is currently in progress on the Geometric Langland applications of boundary line defects \cite{NNpaper}. 

Finally, in this paper we only consider boundary conditions which preserve a full three-dimensional $\CN=4$ supersymmetry algebra, 
including the full three-dimensional Lorentz group. More general half-BPS local boundary conditions which break Lorentz symmetry 
do exist and can play an important role for Geometric Langlands applications.  

\section{Neumann-like boundary conditions and matter interfaces}
Neumann-like half-BPS boundary conditions preserve the full gauge symmetry at the boundary. 
The four-dimensional $G$ gauge fields can be coupled  at the boundary 
to any three-dimensional $\CN=4$ SQFT with $G$ (Higgs branch) global symmetry \cite{Gaiotto:2008sa}.

The boundary conditions for the six scalar fields of SYM gauge theory follow from the boundary conditions on the gauge fields. 
The three scalar fields rotated by $SO(3)_H$ receive Dirichlet boundary conditions: they are set to equal the 
three hyper-K\"ahler moment map operators of the $\CN=4$ SQFT. The three scalar fields rotated by 
$SO(3)_C$ receive Neumann boundary conditions and act as mass parameters for the three-dimensional $\CN=4$ SQFT.

As the system is compactified on $C$, the $\CN=4$ SQFT is twisted as well. The BBB and BAA twists are respectively analogous to the 
Rozansky-Witten (RW) twist \cite{Rozansky:1996bq} and its mirror (mRW). More precisely, the $B$ or $A$ supercharges in complex structure $J$ literally coincide with the 
three-dimensional topological RM or mRW supercharges. This should allow for some interesting tests of our claims using  three-dimensional TFT tools. 

Another simple system one can associate to a $\CN=4$ SQFT with $G$ global symmetry is a ``matter interface'': the four-dimensional gauge fields 
are defined on the whole of space-time and the thee-dimensional $\CN=4$ SQFT is coupled to the restriction of the 
four-dimensional gauge multiplet to a hyper-surface. The three bulk scalar fields rotated by 
$SO(3)_C$ still act as mass parameters for the $\CN=4$ SQFT. The three scalar fields rotated by $SO(3)_H$ are discontinuous at the interface,
with discontinuity given by the three hyper-K\"ahler moment map operators of the $\CN=4$ SQFT.
\footnote{In the language of \cite{Gaiotto:2008sa}, a matter interface is an interface between $G$ and $G$ gauge theories 
with the gauge group restricted to the diagonal $G \subset G \times G$ at the interface.}

\subsection{Pure Neumann boundary conditions}
As a warm-up for the general case, we can review a simple example: pure Neumann boundary conditions without 
any extra boundary degrees of freedom. 

The main subtlety we need to keep track of occurs if the gauge group contains Abelian factors: the Neumann boundary condition 
has a dual Abelian boundary Coulomb branch global symmetry $G_C$ whose currents are simply the boundary values of 
the Abelian bulk field strengths. Concretely, if the gauge group has $n_C$ Abelian factors then $G_C = U(1)^{n_C}$. 

In particular, upon compactification on $C$, configurations involving a gauge bundle of degree $d$ 
give rise to states or Chan-Paton bundles which carry $G_C$ charge $d$. Furthermore, 
the system can be coupled to three-dimensional background twisted $G_C$ vectormultiplets:
a $G_C$ Higgs bundle for a BAA twist and a $G_C$ bundle for a BBB twist. 

The BAA image of pure Neumann boundary condition sets the Higgs field to zero or, more generally, to equal some reference 
one-form $t$ valued in the Abelian part $\mathfrak{g}_A$ of the gauge Lie algebra. 
The gauge connection on $C$ is unconstrained and thus the BAA brane is supported on the full fiber $\CL_t$ of the complex integrable system 
over $t$ in the Hitchin fibration. The definition of an BAA brane also involves a flat connection over $\CL_t$. 
We take that to be a $U(1)$ connection depending only on the Abelian part of the gauge connection, built from a $G_C$ connection 
on $C$. This connection combines with $t$ to give the data of a $G_C$ Higgs bundle. 

As the brane passes through the singular locus of the Hitchin moduli space, this sigma model description of the BAA image of pure Neumann 
boundary conditions may be incomplete. 

The BBB twist of pure Neumann boundary conditions leaves both the connection and the Higgs field free to fluctuate. 
The corresponding BBB brane is thus supported on the whole of $\CM_H(C,G)$. It should be essentially the same as 
the structure sheaf on $\CM_H(C,G)$, up to the possible twist by a point in $\Bun(C,G_C)$, 
interpreted as a choice of line bundle on the torus fibers of the Abelian factors of the Hitchin moduli space. 

The S-dual image of pure Neumann boundary conditions is known in general: it is a modification 
of Dirichlet boundary conditions known as a regular Nahm pole boundary condition. We will review in Appendix \ref{sec:dir} the 
BAA and BBB twists of such boundary conditions. The BAA twist of 
the regular Nahm pole boundary conditions leads to the Hitchin section of the complex integrable system, 
with is the natural mirror to the structure sheaf on $\CM_H(C,G)$. The BBB twist gives a brane whose description
lies outside the scope of the sigma-model description. 

Notice that a Neumann-like boundary condition can be realized as the concatenation of a matter interface and a pure Neumann boundary condition. 

\subsection{Boundary hypermultiplets and interface hypermultiplets}
We now come to the main ingredient of our analysis: boundary hypermultiplets 
valued in some symplectic representation $M$ of $G$. The scalar fields in the hypermultiplet 
transform as a doublet of $SO(3)_H$, while the fermions transform as a doublet of $SO(3)_C$. 

We will denote the complex scalar fields in the hypermultiplets as $Z$, the symplectic pairing on $M$ as $\langle Z, Z' \rangle$
and the complex moment map for the $G$ action on $M$ as $\mu(Z)$. 

In general, the representation $M$ may carry an action of an extra flavor group $G_H$
commuting with the $G$ action. This allows one to couple the system to three-dimensional background $G_H$ vectormultiplets:
a $G_H$ bundle for a BAA twist and a $G_H$ Higgs bundle for a BBB twist. 

The $G_C = U(1)^{n_C}$ symmetry of Neumann boundary conditions is unaffected by the additional matter fields and thus we will 
still be able to couple the system to a $G_C$ Higgs bundle for a BAA twist and a $G_C$ bundle for a BBB twist if the gauge group has Abelian factors.

\subsubsection{BAA twist}
The BAA twist makes the hypermultiplet scalars into spinors on $C$. Thus $Z$ is a spinor section of the symplectic bundle 
$M_E$ associated to $E$, i.e. a section of $M_E \otimes K_C^{1/2}$. Supersymmetry requires the section to be holomorphic.
We can call $\CZ(C,G,M)$ the space of pairs $(E,Z)$.
 
The boundary conditions for the gauge theory scalars require furthermore
\begin{equation} \label{eq:lag}
\Phi = \mu(Z)
\end{equation}
 
Thus the BAA brane should be supported on the image $\CL_M(C,G)$ of $\CZ_M(C,G)$ under the map
\begin{equation}
(E,Z) \to (E, \Phi = \mu(Z))
\end{equation}

It is far from obvious that this image should be a Lagrangian submanifold. It becomes clear if we observe that 
the submanifold manifold $\CL_M$ belongs to a general class of Lagrangian submanifolds associated to generating functions, 
discussed in Appendix \ref{app:BAA}. The generating function for $\CL_M(C,G)$ is the functional 
\begin{equation}
W[Z,\bar A] \equiv \int_C \langle Z, \bar \partial_A Z \rangle
\end{equation}
Indeed the equations defining $\CL_M(C,G)$ can be re-cast in the form 
\begin{equation}
\frac{\delta W}{\delta Z} = \bar \partial_A Z =0 \qquad \qquad \Phi = \frac{\delta W}{\delta \bar A} = \mu(Z)
\end{equation}

In order to specify a BAA brane, we need both the support $\CL_M$ and a choice of flat Chan-Paton bundle. 
The Chan-Paton bundle is given by the bundle of ground states for the system of 
hypermultiplets on $C$ coupled to $E$ and constrained by Equation \ref{eq:lag}.
The system is analogous to a supersymmetric quantum mechanics
with K\"ahler target space and a super-potential $W$ (See Appendix \ref{app:SQM} for details). 
 
If the pre-image of a point under the projection $\CZ_M(C,G) \to \CL_M(C,G)$ is a point or a collection of points, 
they will give a basis for the bundle of ground states. 
If the pre-image of a point $p$ is a manifold $\CF_p$, the flat bundle should be defined 
by the cohomology of $\CF_p$.

If there is a non-trivial flavor group $G_H$, we can couple the system to a $G_H$ bundle on $C$ 
simply by modifying accordingly the $\bar \partial_A$ operator.

Similarly, if $G$ has Abelian factors, we can couple the system to a $G_C$ Higgs bundle 
by deforming Equation \ref{eq:lag} to $\Phi = \mu(Z) + t$ and twisting the Chan-Paton bundle by the 
appropriate line bundle. 

Similar considerations apply for a matter interface consisting of 3d hypermultiplets coupled to the gauge theory at an interface. 
The interface will give a Lagrangian correspondence between two copies of $\CM(C,G)$: 
\begin{equation}
E = E' \qquad \qquad \Phi = \Phi' + \mu(Z)
\end{equation}

\subsubsection{BBB twist}
The BBB twist does not affect the hypermultiplet scalars, but twists the fermions into Grassmann-odd scalars which are 1-forms on $C$. 

The setup is rather analogous to a Nahm transform. The analogy is rather precise, in the sense that both 
the compactification of hypermultiplets on $C$ and all examples of Nahm transforms can be interpreted in the language of 
$N=4$ supersymmetric quantum mechanics, with super-charges described by appropriate Dirac operators. See Appendix \ref{app:SQM}
for details.

Generic enough Higgs field and connection make the bosonic fields and appropriate fermionic partners 
massive, removing them from the calculation. Fermionic zeromodes in the kernel of the Dirac 
operator survive and are quantized to give a fermionic Fock space of supersymmetric ground states. 
\footnote{We are indebted to Kevin Costello for explaining this point to us} 

The Dirac operator is acting on the space of $M$-valued forms on $C$. It is known in the mathematical literature as the Dirac-Higgs operator
\cite{dirachiggs}.  The space of zeromodes of the Dirac-Higgs operator is a hyper-holomorphic sheaf $\CD(C,G,M)$ called the Dirac-Higgs bundle. 
The bundle can be described in complex structure $J$ as the deformed de-Rahm complex of forms valued in $M$ 
with differential $d + \Phi + \bar \Phi$ and in complex structure $I$ as the deformation of the Dolbeault complex by 
$\Phi \wedge$ (``hyper-cohomology''). Generically, the cohomology lies in degree $1$ and gives the Dirac zeromodes. 

The phase space has a (graded) symplectic pairing inherited from the pairing of $M$, which induces a symplectic pairing on the Dirac-Higgs bundle. 
When the bundle lives in degree $1$ we can promote $\CD(C,G,M)$ to a sheaf of Clifford algebras and the sheaf of ground states as the corresponding Clifford module. 

If we are dealing with full hyper-multiplets, i.e. $M = N + N^*$, we can build the corresponding Fock space by using creation operators valued in $N$.
Thus we can take the Dirac-Higgs sheaf $D_N \equiv \CD(C,G,N)$ associated to $N$ and build the Fock space 
\begin{equation}
\CV(C,G,M) \equiv \left( \det D_N\right)^{-1/2} \Lambda^* D_N.
\end{equation}
This is our proposal for the hyper-holomorphic sheaf $\CV(C,G,M)$ defining our BBB brane.

If the Higgs field and connection are not generic enough, the hypermultiplet scalars and their fermionic partners will have zeromodes. 
If the hypermultiplet scalars acquire a supersymmetric vev, they will induce a vev for the bulk scalar fields which 
lie outside the non-linear sigma-model description of the 2d system. We leave a proper understanding of that situation  
to future work. 

The Dirac-Higgs bundle has an intuitive description over loci in the Hitchin moduli space where the determinant of the Higgs field 
in representation $N$ has isolated simple zeroes on $C$: the hypercohomology lies in degree $1$ and decomposes as a direct sum of line bundles, 
the co-kernels for the action of $\Phi$. The corresponding Fock space similarly decomposes into a direct product of 
two-dimensional Fock spaces. If the zeroes of $\Phi$ are isolated but not simple, each multiple zero will contribute an appropriate
vector bundle to the hypercohomology. 

This description is physically reasonable. The Higgs field appears as a mass term in the hypermultiplet Lagrangian. 
We can integrate out the hypermultiplets except at the loci on $C$ where the Higgs field does not have full rank, 
where zeromodes may be localized. We can think about the contributions of each zero to the space of ground states 
as the Hilbert space of a vortex. 

If the Higgs field in representation $N$ vanishes or has non-maximal rank over the whole of $C$, 
these simplifications do not occur and the hypercohomology must be computed more carefully. 

If $M$ includes half-hypermultiplets, i.e. has no Lagrangian splitting, the situation is more subtle. One route may be to double-up the representation
and look for a square root of the Fock space built from $D_M$. Notice that half-hypermultiplets have a potential anomaly, 
which may be reflected into the absence of such a square root. 

For a matter interface, we would place the same sheaf upon the diagonal in $\CM_H(C,G)\times\CM_H(C,G)$.
Thus acting with a matter interface onto a boundary condition simply tensors the corresponding BBB brane by 
the above sheaf $\CB_M$.

If the hypermultiplet representation $M$ also carries a $G_H$ action, we can deform the BBB brane by 
including a solution of $G_H$ Hitchin equations into the Dirac-Higgs operator. In particular, the hypercohomology will be supported 
on loci in $C$ where $\Phi + \Phi_H$ drops in rank. 

In Appendix  \ref{app:SQM} we will propose an alternative description of the Dirac-Higgs sheaf involving BAA branes in $T^*C$. 

\subsection{Boundary gauge theories}
A more general class of boundary conditions may be obtained by coupling the boundary hypermultiplets 
to some $G_B$ boundary gauge fields.

We can describe this system as the composition of an interface with boundary hypermultiplets 
coupled to $G$ and $G_B$ four-dimensional gauge fields and a $G_B$ pure Neumann boundary condition.

The BAA and BBB twisted descriptions of the system should analogously follow from the composition of the appropriate 
2d interfaces and boundary conditions. 

\subsubsection{BAA twist}
We will now consider an auxiliary space $\CZ(C,G, G_B,M)$ consisting of triples $(E,E_B,Z)$,
with $Z$ coupled to both the $G$ bundle $E$ and the $G_B$ bundle $E_B$, subject to the further constraint 
\begin{equation}
\mu_B(Z) = 0
\end{equation}
where $\mu_B(Z)$ is the moment map for the $G_B$ action. If $G_B$ has Abelian factors, the right hand side can be shifted by a 1-form. 

The BAA brane should be again supported on the image $\CL(C,G,G_B,M)$ of $\CZ_M(C,G,G_B,M)$ under the map
\begin{equation}
(E,E_B,Z) \to (E, \Phi = \mu(Z))
\end{equation}

\subsubsection{BBB twist}
The analysis of the previous section gave us a hyper-holomorphic bundle, or sheaf,
$\CV(C,G,M)$ on $\CM_H(C,G) \times \CM_H(C,G_B)$. In order to understand the boundary conditions involving a
$G_B$ gauge theory, we need to convolve that sheaf with the structure sheaf on $\CM_H(C,G_B)$. 

In a given complex structure, the B-model prescription for the convolution is 
that the stalks of the resulting sheaf at a point on $\CM_H(C,G)$ are the global sections or, better, the
cohomology of $\CV(C,G,M)$ restricted to $p$. 

Hyper-holomorphically, the resulting sheaf should be the sheaf of supersymmetric ground states of an $\CN=4$ SQM with $\CM_H(C,G_B)$ target space, associated to the hyper-holomorphic connection on $\CB_M$. See Appendix \ref{app:SQM} for details. This is roughy a
Nahm transform of $\CV(C,G,M)$ over $\CM_H(C,G_B)$. 

It is reasonable to worry, though, that this naive description may miss contributions from the singular loci of $\CM_H(C,G_B)$,
where unbroken $G_B$ gauge fields are present. A description as an $\CN=4$ gauged SQM may be more suitable. 
We leave this problem to future work. 

\subsection{Generalization to smooth sigma models}
Our discussion of the free hypermultiplet case does generalize in a natural way to boundary conditions involving 
three-dimensional non-linear sigma model. The sigma models must have a smooth hyper-K\"ahler target space $\CX$
equipped with a tri-holomorhic $G$ action. 

The Higgs branch global symmetry $G_H$ of such a model consists of extra tri-holomorhic isometries commuting with $G$.
If $\CX$ is endowed with suitable $(1,1)$ forms, such as these associated with resolution/deformation parameters, they can be pulled back to space-time 
to define currents for Abelian $G_C$ symmetries. The mass parameters for such isometries are identified with the resolution/deformation parameters. 
Notice that the sigma model description will break down if these parameters are tuned to a value which makes $\CX$ singular. 

The BAA twist of such a boundary condition requires $\CX$ to admit a $SO(2)_H$ K\"ahler isometry in complex structure $I$ which rotates the complex symplectic form,
realizing the Cartan of $SO(3)_H$.  The $SO(2)_H \times G$ action can be used to make $\CX$ into the fiber of a bundle $\CX_C$ on $C$, 
in such a way that the complex symplectic form on $\CX$ is a section of $K_C$ and the complex $G$ moment map a section of $\End(E) \otimes K_C$. 
Then it makes sense to consider holomorphic sections $Z$ of $\CX_C$ and write $\Phi = \mu(Z)$ to define a BAA brane. 

Notice that the complex deformation parameters $t_\C$ of $\CX$ would break $SO(2)_H$. They can be introduced in the story 
if they are twisted to a 1-form $t(z)$ on $C$. This is the background Higgs field for the $G_C$ twisted vectormultiplet. 

In the BBB twist, we can consider a generalization of the $(0,4)$ supersymmetric quantum mechanics of free hypermultiplets on $C$ discussed in Appendix \ref{app:SQM}.
We can take $(0,4)$ hypermultiplets valued in $\CX$ together with a bundle of $(0,4)$ Fermi multiplets modelled on 1-forms on $C$ valued in the pull-back of the tangent bundle 
to $\CX$. 

An example of a situation where the non-linear sigma model construction is fully applicable involves a family of hyper-K\"ahler manifolds 
defined as moduli spaces of solutions of Nahm equations on a segment with structure group $G$. When the boundary conditions 
are selected to be Dirichlet at one end of the segment, a Nahm pole on the other end, the result is a smooth hyper-K\"ahler manifold with $G \times Z_\rho$
hyperholomorphic isometries, where $Z_\rho$ is the stabilizer of the Nahm pole (see \cite{Gaiotto:2008sa} for a review and several mathematical references).
Modified Neumann boundary conditions involving these manifolds are equivalent to the Dirichlet and Nahm pole boundary conditions reviewed in 
Appendix \ref{sec:dir}. They provide an alternative way to understand the image of these boundary conditions upon compactification on $C$. 
 
An important potential application of non-linear sigma model construction would be to approximate the theory $T[G]$ by its resolved Higgs branch, 
the co-tangent bundle to the flag variety for $G$. This approximation would not be compatible with the most general possible 
background for the ${}^\vee G$ valued twisted vectormultiplets, but should be sufficient to deal with 
backgrounds in an Abelian subgroup of ${}^\vee G$. It would be interesting to match this construction with 
the mathematical construction of the Geometric Langland dual to local systems in an Abelian subgroup of ${}^\vee G$.
See e.g. \cite{A.Braverman:aa}.

\section{A rich example: particle-vortex duality in $U(1)$ gauge theory}
The $U(1)$ Hitchin equations are linear and thus the corresponding moduli space is rather simple:
it is the product of the space of 1-forms $\Phi$ on $C$ and of the space $\Jac(C)$ of degree $0$ line bundles $E$ on $C$. 

There are two mirror subtleties which are related to the existence of an unbroken four-dimensional gauge symmetry. 
Boundary conditions can carry magnetic and electric charges $d$ and $q$
which are roughly exchanged by S-duality. 
\begin{itemize}
\item The magnetic charge modifies the Hitchin moduli space by shifting the degree of the line bundle from $0$ to $d$ so that the 
fiber of the Hitchin fibration is $\Pic_d(C)$
We obtain an infinite family of moduli spaces $\CM_d$, labelled by the degree $d$ of the line bundle $E$.
Giving a boundary condition means giving a brane $B_d$ in each of these moduli spaces. 
\item There is an unbroken 2d gauge symmetry and thus the Chan-Paton bundles should really be thought of as 
a choice of boundary condition for that 2d gauge theory. Essentially, they should carry an action of the unbroken gauge group. 
Boundary conditions can be decomposed into branes labelled by such extra representation data $q$. Geometrically, this is 
recognizable as an ambiguity in the Chan-Paton bundle which can be fixed non-canonically by picking a specific 
representative for $E$. 
\end{itemize}

These subtleties are discussed at length in \cite{Witten:2015aa} as they affect the treatment of Wilson and 't Hooft lines.
A 't Hooft line maps to a BAA interface which identifies the Higgs fields $\Phi$ and $\Phi'$ on the two sides, but relates
the line bundles $E$ and $E'$ by a Hecke modification at a point, which shifts the degree of the bundle. 
A Wilson line maps to a BBB brane on the diagonal of the two Hitchin moduli spaces, whose 
Chan-Paton bundle $E_z$ is given by the line bundle restricted to a point $z$ on $C$. Different choices 
representative for $E$ over the space of line bundles give a different meaning to the Chan-Paton bundle.

At Neumann boundary conditions, there is a ``topological'' $U(1)_t$ factor in $G_C$ which is generated by the current $*F$.
The degree $d$ can be identified with the $U(1)_t$ charge of the state. In the presence of matter fields, there is a slightly spurious, but still useful, 
$U(1)_b$ factor in $G_H$ which acts on the boundary matter fields in the same way as the bulk gauge group. The charge $q$ can be identified with the 
$U(1)_b$ charge of the state. 

The simplest non-trivial example of the action of S-duality on boundary conditions 
arises for a $U(1)$ gauge theory coupled at the boundary to a single hypermultiplet of charge $1$. 
This boundary condition is self-mirror. It has both $G_H$ and $G_C$ equal to $U(1)$. 

\subsection{BAA analysis}
We decompose the hypermultiplet scalars into two complex fields $X$ and $Y$ of gauge charge $1$ and $-1$ respectively. 

Given a $U(1)$ bundle $E$, $X$ is a section of $E \otimes K_C^{1/2}$ and $Y$ is a section of $E^{-1} \otimes K_C^{1/2}$.
The Higgs field is then $\Phi = X Y$. Notice that the difference in the dimension of the spaces of $X$ and $Y$ global sections 
is controlled by an index theorem and equals the degree $d$ of $E$.  

At first, we can take the degree of $E$ to be $0$. Generically there are no holomorphic $X$ and $Y$ and we have $\Phi=0$. 
At a special co-dimension 1 locus on $\Bun_0(C,U(1))$, the theta divisor $\Theta_0$, holomorphic solutions for $X$ and $Y$ will occur 
simultaneously. 

At generic points on $\Theta_0$, there is a one-dimensional space of solutions for $X$ and $Y$ and thus a 1-dimensional space 
of possible $\Phi$, which parameterizes the co-normal bundle to $\Theta_0$. Above a point in that Lagrangian image we have a $\C^*$ 
worth of possible pairs $(X,Y)$, acted upon by the unbroken 2d gauge symmetry. At special loci on the $\Theta_0$ divisor the dimension of the space 
of solutions for $X$ and $Y$ may grow larger and the local geometry of the Lagrangian manifold will become more intricate and possibly singular. 

In positive degree $d$, there will generically be a $d$-dimensional space of solutions for $X$ and none for $Y$. 
We have again $\Phi=0$ but a non-trivial $\C^d$ fiber acted upon by the unbroken gauge transformations. 
At a co-dimension $d+1$ locus $\Theta_d$ in the space of bundles a solution for $Y$ will appear and there will be a $(d+1)$-dimensional space of solutions for $X$. 
Now $\Phi$ will also live in a $(d+1)$-dimensional space. More complicated geometry will again occur in higher co-dimension. 

For $d<0$ the roles of $X$ and $Y$ are exchanged. 

In order to study fiberwise mirror symmetry, we need to understand the intersection of the Lagrangian manifold with the fibres of the Hitchin fibration. 
A non-zero Higgs field $\Phi$ has $2g-2$ zeroes. For simplicity, we can assume that they are distinct. The zeroes of $\Phi$ must be either zeroes of $X$ or $Y$: 
$g-1+d$ zeroes of $X$ and $g-1-d$ zeroes of $Y$. 

Conversely, we can find a solution for any of the ${2g-2 \choose g-1+d}$ possible ways of splitting the zeroes of $\Phi$ among $X$ and $Y$: given the zeroes, 
the bundle $E$ exists and is unique. We can roughy represent it by a divisor $\frac{\sum_i x_i - \sum_j y_j}{2}$, where $x_i$ and $y_j$ are the zeroes in $X$ and $Y$,
by looking at the meromorphic section $X/Y$ of $E^2$. 

Thus the Lagrangian manifold is a nice multi-section of the Hitchin fibration, with ${2g-2 \choose g-1+d}$ sheets, ramified above the locus where $\Phi$ has
higher order zeroes. Above a generic point in the manifold there is a
$\C^*$ worth of possible values of $X$ and $Y$, acted upon by the unbroken 2d gauge symmetry. 

Above the singular point $\Phi=0$, instead, we have the whole fiber of the Hitchin fibration.

The mirror of this BAA brane should be a collection of sheaves which at generic $\Phi$ have rank  ${2g-2 \choose g-1+d}$ and when restricted to the fiber at that point 
is a sum of line bundles of the form $\left( \otimes_i L_{x_i}^{\frac{1}{2}} \right) \otimes \left( \otimes_j L_{y_j}^{-\frac{1}{2}} \right)$, where $L_z$ is the line bundle on the fiber which 
associates to an element $L$ of $\Pic_d$ the restriction of $L$ to the point $z$. We refer to \cite{Witten:2015aa} for details of the Abelian fiberwise mirror map.

\subsection{BBB analysis} 
If the zeroes of $\Phi$ are distinct and $E$ has degree $0$, the hyper-cohomology of $(E,\Phi)$ in a charge $1$ representation 
consists of a direct sum 
\begin{equation}
{\bf H}^1(E,\Phi) = \oplus_{i=1}^{2g-2} \left(E \otimes K\right)_{z_i}
\end{equation}
where $z_i$ are the zeroes of $\Phi$. and the summands on the right hand side are the restriction of $E \otimes K$ to these points.

Correspondingly, the Fock space is 
\begin{equation}
F{\bf H}^1(E,\Phi) = \otimes_{i=1}^{2g-2} ((E\otimes K)_{z_i}^{1/2} \oplus (E \otimes K)_{z_i}^{-1/2})
\end{equation}
The two summands in each product have $G_H$ charges $\pm 1/2$ respectively. 

Thus the BBB brane, for non-trivial $\Phi$ with distinct zeroes, looks like a direct sum of rank ${2g-2 \choose g-1+q}$ vector bundles.
The dimension of this space, including the $G_H$ grading, matches the number of points in the fiber 
above $\Phi$ in the conjectural mirror BAA brane, graded by degree. 

Each summand of the Fock space can be written as $\otimes_i (E\otimes K)_{x_i}^{1/2}\otimes_j (E\otimes K)_{y_j}^{-1/2}$,
where we split the zeroes of $\Phi$ into the two groups $x_i$ and $y_j$. 
This line bundle is naturally mirror to the divisor $\frac{\sum x_i - \sum y_i}{2}$, matching the BAA analysis.
Indeed $L$ there is the same as $E$ here and the extra factor of $K$ does not change the resulting bundle 
on the fiber of the Hitchin fibration. 

We expect a similar analysis to hold for degree $d$ bundles, giving again a sum of terms of the form
$\otimes_i (E\otimes K)_{x_i}^{1/2}\otimes_j (E\otimes K)_{y_j}^{-1/2}$. The existence of this infinite collection of branes of the same form for various $d$ 
should be mirror to the presence of the $\C^*$ boundary degrees of freedom in the BAA analysis.

In conclusion, we verified that at least above a generic point in the base of the integrable system the BBB and BAA images of the four-dimensional boundary condition 
are mirror to each other. 

\section{Bifundamental and fundamental interfaces}
A simple modification of the setup in the last section is to take a matter interface for a $U(1)$ gauge theory consisting of a 
single charge $1$ hypermultiplet. Applying S-duality to both sides of the interface, we end up with a 
``bi-fundalental'' interface: Neumann boundary conditions for both $U(1)$ gauge groups, 
with a hypermultiplet of charge $(1,-1)$ at the interface. 

Although this system is closely related to the one in the previous section, it has a nice advantage:
it can be generalized to $U(N)$ gauge theory. S-duality relates a matter interface with hypers in a fundamental representation  for a $U(N)$ gauge theory (``D5 interface'')
to an interface involving bi-fundamental hypermultiplets coupled to two $U(N)$ gauge theories with Neumann b.c. (``NS5 interface''). 
We will now consider such a generalization.

\subsection{BAA analysis of a D5 interface}
As this is a matter interface and the gauge fields continue across the interface, the gauge bundles on the two sides of the interface coincide, i.e. we have $E=E'$.
On the other hand, the Higgs fields satisfy
\begin{equation} \label{eq:dbaa}
\Phi' = \Phi + X Y
\end{equation} 
where $X$ and $Y$ are sections of $E \otimes K_C^{1/2}$ and $E^* \otimes K_C^{1/2}$ respectively. 

Consider a generic location on the bases of the Hitchin moduli spaces where we can encode the Higgs bundles into 
spectral curves $\Sigma$ and $\Sigma'$ in $T^*C$. In the notation of \cite{Witten:2015aa}, the spectral data of the Higgs 
bundle $(\Phi, E)$ is the spectral curve $\det (y - \Phi)=0$ in $T^*C$ equipped with a line bundle 
$L$ such that $E$ is the pushforward of $L$ along the projection to $C$. 

Over the spectral curve $\Sigma$ we have both the 1-form $\lambda$ of $\Phi$ eigenvalues and the corresponding eigenvector $v$, which can be thought of as 
a section of the pull-back of $E$ to the spectral curve $\Sigma$ tensored with the line bundle $\tilde L$ whose pushforward to $C$ is $E^*$. Over $\Sigma'$ we have both 
the 1-form $\lambda'$ of $\Phi'$ eigenvalues and the corresponding left eigenvector $v'$, section of the pull-back 
of $E^*$ to $\Sigma'$ tensored with the line bundle $L'$ whose pushforward to $C$ is $E'$ .

Pulling back the above equation to the combined curve parameterized by the pair $(\lambda, \lambda')$ over $C$
and contracting with the eigenvectors, we find 
\begin{equation}
(\lambda' - \lambda) (v',v) = (v',X)(Y,v)
\end{equation}

The zeroes of $\lambda-\lambda'$ are zeroes of a polynomial in the traces of $\Phi$ and $\Phi'$, the {\it resultant} of their characteristic polynomials. 
It is a differential of degree $N^2$ on $C$. These zeroes will have to match either a zero of $(v',X)$ on the point of $\Sigma'$ labelled by $\lambda'$ or 
of $(Y,v)$ on the point of $\Sigma$ labelled by $\lambda$. 

These two inner products are sections of the pull-back of $K^{\frac12}$ to $\Sigma'$ or $\Sigma$ tensored respectively with $L'$ and $\tilde L$. 
The degree of these bundles is $N^2(g-1) \pm d$ and thus the total number of zeroes they have is $n = N^2 (2g-2)$, 
which is the same as the total number of zeroes of the resultant. 

Thus if the resultant has simple zeroes, in order to find all solutions of \ref{eq:dbaa}
we can pick any of the $2^n$ ways to split the zeroes of the resultant among $(v',X)$ and $(Y,v)$ and then characterize $\tilde L$ and $L'$ 
by the corresponding divisors. Each of these possible $2^n$ choices will fix the line bundle on the spectral curves and determine an intersection point 
of our Lagrangian manifold with the fibers of the complex integrable systems. 

Upon fiberwise mirror symmetry, this BAA brane will map to a sheaf with rank $2^n$ over the product of the two Hitchin systems, 
which above generic points on the base of the integrable systems
will be described as a sum of line bundles which are mirror dual to the product of the $L$ and $L'$ associated to the possible splittings of the zeroes of 
the resultant.

\subsection{BBB analysis of a D5 interface}
The BBB twist of the interface will provide a sheaf supported on 
the diagonal of the product of the two $U(N)$ Hitchin systems associated to the gauge theories on either sides of the matter interface.
 
The sheaf is the Fock space built from the fundamental Dirac-Higgs bundle for the Hitchin system. 
It has rank $2^{2N(g-1)}$.

At least over loci where $\det \Phi$ has simple zeroes, the Dirac-Higgs bundle 
is again a direct sum of sub-bundles, the co-kernels of the map 
$E_{z_i} \to (E\otimes K)_{z_i}$ given by the action of $\Phi(z_i)$, 
where $z_i$ are the zeroes of $\det \Phi$. The Fock space is built as before.

The co-kernels can be recognized as the restriction of the line bundle 
$L \otimes K_C$ on the spectral curve to the points where the spectral curve intersects the zero section of 
$T^* C$. The Fock space will be a direct sum of terms of the rough form 
\begin{equation}
\left( \otimes_i (L \otimes K_C)_{x_i}^{\frac12} \right) \otimes \left( \otimes_j (L \otimes K_C)_{y_j}^{-\frac12} \right)
\end{equation}
where $x_i$ and $y_j$ are a partition of the zeroes of $\det \Phi$ in two groups.
See Appendix \ref{app:SQM} for an A-model version of this statement. 

\subsection{BAA analysis of an NS5 interface}
The Higgs fields for the two gauge theories satisfy now
\begin{equation}
\Phi' = X Y \qquad \qquad \Phi = Y X
\end{equation} 
where $X$ and $Y$ are sections of $E' \otimes E^* \otimes K_C^{1/2}$ and $E \otimes (E')^* \otimes K_C^{1/2}$ respectively. 

At least as long as $\Phi$ and $\Phi'$ have full rank, the two Higgs fields have the same characteristic polynomial.
Taking the determinant of the relations, we get that over loci where $\det \Phi$ has simple zeroes, the $2 N (g-1)$ zeroes of $\det \Phi$ must be distributed among 
$\det X$ and $\det Y$. 

Furthermore $X$ and $Y$ map eigenvectors of $\Phi$ into eigenvectors of $\Phi'$ 
with the same eigenvalues and viceversa. They provide an identification of the pair of bundles $L$ and $K^{\frac12} \otimes L'$ or the pair of bundles $L'$ and $K^{\frac12} \otimes L$ 
on the spectral curve away from the zeroes of $\lambda$. Thus $L$ and $K_C^{\frac12} \otimes L'$ or $L'$ and $K_C^{\frac12} \otimes L$ can only differ by a modification 
at these zeroes. As $\lambda$ has $2N (g-1)$ zeroes and $K_C^{\frac12}$ has degree $N(g-1)$, it is reasonable to 
conclude that 
\begin{align}
K_C^{\frac12} \otimes L' &= L \otimes \left( \otimes_i O(x_i) \right) \cr
K_C^{\frac12} \otimes L &= L' \otimes \left( \otimes_j O(y_j) \right)
\end{align}
where $x_i$ and $y_j$ are a partition of the zeroes of $\det \Phi$ in two groups.

We can confirm the linear relation between $L$ and $L'$ by observing that 
\begin{equation}
\Phi' X = X \Phi  \qquad \qquad Y \Phi' = \Phi Y
\end{equation} 

These relations imply directly that the Lagrangian submanifold is fixed by the simultaneous 
Hamiltonian flow along the fibers of the two complex integrable systems. Indeed, the flow is generated by shifting 
$\bar \partial_E$ by an amount proportional to a polynomial the Higgs field. Then equations such as $\bar \partial_{E\times (E')^*} X=0$ 
change along the flow by amounts such as $\Phi' X - X \Phi$, which vanish on the Lagrangian submanifold.

Thus these intertwining relationships show that 
the Lagrangian submanifold is invariant under differences of flows in pairs of copies,
generated by corresponding pairs  $u- u'$ of Hamiltonians in the two systems. 
In terms of angular coordinates $\theta$ and $\theta'$ on generic  
fibers, the Lagrangian should be defined locally by equations of the form $u = u'$ and $\theta'-\theta=f(u)$.

This is a rather reasonable property for a Lagrangian correspondence which sits over the 
diagonal on the bases of the complex integrable system: the fiber consists of $2^{2 N(g-1)}$ shifted images of the diagonal,
labelled by the splitting of zeroes of $\det \Phi$ among $x_i$ and $y_j$.

Upon mirror symmetry, we should get a BBB brane 
associated to a rank $2^{2 N(g-1)}$ sheaf on the diagonal. This agrees with the prediction of S-duality
between NS5 and D5 interfaces. Indeed, above general points in the base we can 
match the difference between $L$ and $L'$  for each summand of the sheaf 
to the result of the BBB analysis for the D5 interface. This agrees with the prediction of S-duality
between NS5 and D5 interfaces. 

\subsection{BBB analysis of an NS5 interface}
The BBB twist of the interface will provide a sheaf on the product of the two rank $N$ Hitchin systems. 
The sheaf is the Fock space built from the Dirac-Higgs bundle in the bi-fundamental representation over the 
product of Hitchin system. 

If the rank of the linear map $f \to \Phi f - f \Phi'$ is maximal away from isolated points on $C$. 
the Dirac-Higgs bundle should decompose again into sub-bundles, the co-kernels of that 
linear map. These $n$ points are the zeroes of the resultant of the characteristic polynomials of $\Phi$ and $\Phi'$. 

The co-kernel consists essentially of the tensor product of the $\Phi$ eigenvector and $\Phi'$ left eigenvector 
with matching eigenvalues. Thus the Dirac-Higgs bundle can be expressed in terms of the product of the 
line bundles on the spectral curves at the points where they intersect. See Appendix \ref{app:SQM} for an A-model version of this statement.

At such generic loci, the Fock bundle has thus rank $2^n$ and is a sum of line bundles which can be matched under fiberwise mirror symmetry 
with the results of the BAA analysis for the D5 interface. This agrees with the prediction of S-duality
between NS5 and D5 interfaces. 

\section{General NS5 and D5 interfaces for unitary groups}
The D5 and NS5 interfaces have a generalization to interfaces between $U(N)$ and $U(M)$ gauge theories
with $M<N$. On the NS5 side the generalization is obvious: the interface supports $N \times M$ hypermultiplets 
in a bi-fundamental representation of $U(N) \times U(M)$. On the D5 side, one has no matter, but an interface with reduced 
gauge symmetry. At the interface, $U(M)$ is identified with a block-diagonal subgroup of $U(N)$ and a Nahm pole is inserted 
in the remaining $U(N-M)$ block of $U(N)$.

Due to the Nahm pole, the BBB analysis of the D5 interface goes outside of the sigma model description: 
the BBB brane is supported on the singular locus in the product of Hitchin systems 
where the solution of $U(N)$ Hitchin system reduces to a solution of the $U(M)$ Hitchin system times a trivial 
solution in of the $U(N-M)$ Hitchin system. 

Within limits, we can reproduce this from the BAA twist of the NS5 interface. 
The equations 
\begin{equation}
\Phi' = X Y \qquad \qquad \Phi = Y X
\end{equation} 
now tell us that the $U(N)$ Higgs field $\Phi$ has rank at most $M$, with $X$ and $Y$ 
providing the inter-twiners between the $U(N)$ and $U(M)$ Higgs bundles. 

More precisely, if $\Phi'$ has full rank, then $\Phi$ must have rank $M$ and the characteristic polynomial 
of $\Phi$ should be $ \det (x - \Phi) x^{N-M} \det (x - \Phi')$. 

The relations
\begin{equation}
\Phi' X = X \Phi  \qquad \qquad Y \Phi' = \Phi Y
\end{equation} 
still hold, showing that the Lagrangian manifold is fixed by the simultaneous 
Hamiltonian flow along the fibers of the two complex integrable systems

The other twists are more interesting.

\subsection{BBB twist of NS5 interface}
The Dirac Higgs bundle is again built from the cokernels of the map 
$f \to \Phi f - f \Phi'$. The rank of that map drops at the $n= N M (2g-2)$ points 
which are the zeroes of the resultant of the characteristic polynomials of $\Phi$ and $\Phi'$. 

Again, if the zeroes are simple, the cokernels can be identified with matrices which are the tensor product of the 
eigenvectors of $\Phi$ and $\Phi'$ with a common eigenvalue.  

At such generic loci on the bases of the Hitchin systems, the Fock bundle has thus rank $2^n$ and is a sum 
of line bundles built from the cokernels in the usual fashion. 

\subsection{BAA twist of D5 interface, $M=N-1$}
It is instructive to specialize at first to $N-M=1$. The boundary condition imposes (see Appendix \ref{sec:dir})
\begin{equation}
E = E' \oplus L
\end{equation}
for some fixed line bundle $L$. Here $L$ gives the coupling to a background 
vectormultiplet for the $U(1)$ global symmetry of the interface. \footnote{This could also have been included in the 
$N=M$ case by twisting the bundle $X$ and $Y$ transformed in.}

We also have the constraint 
\begin{equation}
\Phi|_{\End(E')} = \Phi'
\end{equation}
Pulling back this relation to the (fiberwise product of the) spectral curves and sandwiching between eigenvectors $v = (v_{E'},v_L)$ for $\Phi$ and 
$(v',0)$ with $v'$ eigenvector for $\Phi'$ we get
\begin{equation}
(\lambda' - \lambda) (v_{E'}, v') = v_L (\Phi|_{\Hom(E',L)} v')
\end{equation}
The zeroes of $\lambda' - \lambda$ are zeroes of the resultant of the two characteristic 
polynomials of the Higgs fields, which is a differential of degree $NM$. Thus it has 
$n = N(N-1)(2g-2)$ zeroes. 

Again, the $n$ zeroes of $\lambda' - \lambda$ will be distributed among the 
sections $v_L$ and $(\Phi|_{\Hom(E',L)} v')$ on the two spectral curves. 
We expect each of these possible $2^n$ choices to determine the line bundle on the spectral curve and thus an intersection point 
of the Lagrangian manifold with the fibers of the complex integrable systems. 

With a bit more work, it should be straigthforward to fully verify the fiberwise mirror symmetry at generic fibers 
which follows from the prediction of S-duality between NS5 and D5 interfaces. 

\subsection{BAA twist of D5 interface, general $N-M$}
Because of the Nahm pole (see Appendix \ref{sec:dir}), the bundles on the two sides are related as 
\begin{equation}
E = E' \oplus \left( L \otimes K^{\frac{N-M-1}{2}} \right) \oplus \cdots \oplus \left( L \otimes  K^{\frac{-N+M+1}{2}} \right)
\end{equation}
with $\Phi|_{\End(E')} = \Phi'$. 

We can use gauge transformations to restrict the form of $\Phi$ further, analogously to what happens for the 
Hitchin section at $M=0$. This can be encoded in terms of the regular $su(2)$ embedding in $U(N-M)$.
The canonical form for $\Phi$ is that of a Slodowy slice, the sum of a raising operator $t^+$ and a piece which commutes with the lowering operator 
$t^-$. In particular, the off-diagonal blocks of $\Phi$ reduce to a single column or row with $M$ elements each. 

Now we can sandwich $\Phi$ between eigenvectors for $\Phi$ and $\Phi'$ as before, 
to get the usual equation 
\begin{equation}
(\lambda' - \lambda) (v_{E'}, v') = v_{L \otimes K^{\frac{N-M-1}{2}}} (\Phi|_{\Hom(E',L \otimes K^{\frac{N-M-1}{2}})} v')
\end{equation}
and proceed as before to match the zeroes of $\lambda' - \lambda$ with the zeroes of the two factors on the right hand side. 

\subsection{Hanany-Witten moves}
Hanany-Witten moves play a crucial role in understand the S-duality properties of boundary conditions in $U(N)$ gauge theory.

The basic idea is that boundary conditions defined by linear quiver gauge theories can be built as a sequence of 
D5 and NS5 interfaces. The NS5 interfaces provide the bi-fundamental hypermultiplets 
and the D5 interfaces can provide extra fundamental hypermultiplets. 

S-duality is implemented by mapping D5 to NS5 and viceversa. After this map, it is often useful to change the relative order 
of D5 and NS5s in such a way to eliminate D5 interfaces involving Nahm poles or symmetry breaking. This can be done with the help of the 
Hanany-Witten moves. It would be very useful to prove directly the relations between convolutions of D5 and NS5 interfaces which follow from the 
Hanany-Witten rules: it would allow one to repeat the brane manipulations directly in the language of 
Geometric Langlands. We leave this problem for future work.  

\section{Ortho-symplectic examples}
Brane constructions extend to orthogonal and symplectic gauge groups, albeit with 
several non-trivial complications. 

A simple generalization of NS5 interfaces for unitary groups are the half-NS5 interfaces 
between an orthogonal group $SO(2N)$ and a symplectic group $USp(2M)$. They consist simply of a
bi-fundamental hypermultiplet, i.e. a hypermultiplet in the product of the fundamental representations, which is obviously 
symplectic. 

The S-dual of these interfaces are half-D5 interfaces between the orthogonal group $SO(2N)$ and the orthogonal group $SO(2M+1)$.
They are defined by a Nahm pole of dimension $|2N-2M-1|$ together with an embedding of the smaller group into the larger. 

It is also possible to consider half-NS5 interfaces between an orthogonal group $SO(2N+1)$ and a symplectic group $USp(2M)$,
defined in an analogous manner. Notice that this interface involves an odd number 
of half-hypermultiplets for $USp(2M)$, which have an anomaly. The anomaly is compensated 
by a discrete theta angle in the $USp(2M)$ gauge theory, which we can thus denote as $USp(2M)'$. 

The theta angle affects the S-duality properties of the $USp(2M)'$ gauge theory, which is mapped back to 
$USp(2M)'$. The S-dual interface is expected to be a half-D5 interface between $USp(2N)$ and $USp(2M)'$.
If $N=M$, this involves a fundamental half-hypermultiplet. Otherwise, it involves 
the embedding of the smaller gauge group into the larger and a Nahm pole of dimension 
$2|N-M|$. 

It would be very interesting to work out these general examples in full detail. Instead, in the next sections we will limit ourselves to 
discussing some simple examples involving $SU(2)$ gauge groups. For example, the half-NS5 interface between 
$SO(2)$ and and $USp(2)$ will be discussed as an interface between $U(1)$ and $SU(2)$. This is dual to the half-D5 interface 
between $SO(2)$ and $SO(3)$. Similarly, the half-NS5 interface between $SO(1)$ and $USp(2)'$ consists of a single half-hypermultiplet 
for a single $SU(2)$ gauge theory and is dual to the half-D5 interface between $USp(0)$ and $USp(2)'$, i.e. the maximal  Nahm 
pole for $SU(2)'$. Finally, the half-NS5 interface between $SO(4)$ and $USp(2)$ is really a tri-valent interface 
between three $SU(2)$ gauge groups with very nice properties, S-dual to the half-D5 interface between $SO(4)$ and $SO(3)$,
which identifies the three $SU(2)$ gauge groups with each other. 

\section{Tri-fundamental $SU(2) \times SU(2) \times SU(2)$ interface}
Until now we have considered either boundary conditions for a single four-dimensional gauge theory, 
leading to branes in a single Hitchin moduli space, or interfaces between two four-dimensional gauge theories, leading to 
branes in the product of two Hitchin moduli spaces, possibly interpreted as correspondences or functors. 

In this section we look at an example of interface between three gauge theories, leading to 
branes in the product of three Hitchin moduli spaces. This could also be used to produce a
functorial operation on the category of branes. The particular example we consider here actually gives an associative 
operation. It would be nice to explore it further. 

The simplest possible tri-valent interface is a ``tri-transparent'' interface between three $G$ gauge theories, where the three gauge groups 
are simply identified at the interface. In other words, we preserve at the interface the diagonal combination of the three 
$G$ gauge groups, without adding any extra matter fields. 

For $G=SU(2)$, the interface is S-dual to a Neumann boundary condition for all three gauge groups coupled to 
a single non-anomalous half-hypermultiplet transforming in 
the product of the fundamental representations of $SU(2) \times SU(2) \times SU(2)$.

This fact follows from the properties of half-NS5 and half-D5 interfaces for orthogonal and symplectic gauge theories, but can also be derived 
with a bit of help of the 6d $(2,0)$ theory. The generalization to other groups $G$ involves a non-trivial interface SCFT 
which brings the problem out of the scope of this paper. 

We will thus analyze this three-way interface for $SU(2) \times SU(2) \times SU(2)$
and compare it to the behaviour of the ``tri-transparent'' interface.

\subsection{BAA analysis of tri-fundamental}
A mathematical discussion of this system was recently given in \cite{Hitchin:2016aa}.

The hypermultiplet scalar field $Z$ is a section of $E \otimes E' \otimes E'' \otimes K^{1/2}$,
where $E$, $E'$, $E''$ are the bundles associated to the irreducible doublet representations of the three $SU(2)$ groups. We will explicitly denote the $8$ components as
$Z_{\alpha \beta \gamma}$, where the indices transform under the three $SU(2)$. 

The corresponding Lagrangian manifold in the product of three copies of the $SU(2)$ Hitchin system is defined by the equations
\begin{align}
D_{\bar z} Z &=0\cr
\Phi^{(a)} &= \mu^{(a)}(Z)
\end{align}
there $\mu^{(a)}(Z)$ are the moment maps for the three $SU(2)$ groups and $\Phi^{(a)}$ the corresponding Higgs fields.
We will sometimes denote the three Higgs fields as $\Phi$, $\Phi'$, $\Phi''$. 

In components, we can write the equations as 
\begin{align}
\Phi^{(1)}_{\alpha \alpha'} &= \frac{1}{2} Z_{\alpha \beta \gamma} Z_{\alpha' \beta' \gamma'} \epsilon^{\beta \beta'} \epsilon^{\gamma \gamma'}\cr
\Phi^{(2)}_{\beta \beta'} &= \frac{1}{2} Z_{\alpha \beta \gamma} Z_{\alpha' \beta' \gamma'} \epsilon^{\alpha \alpha'} \epsilon^{\gamma \gamma'}\cr
\Phi^{(3)}_{\gamma \gamma'} &= \frac{1}{2} Z_{\alpha \beta \gamma} Z_{\alpha' \beta' \gamma'}  \epsilon^{\alpha \alpha'} \epsilon^{\beta \beta'}
\end{align}
where the $\epsilon$ symbols denote the symplectic pairing in each doublet ($\epsilon^{+-}=1$, $\epsilon^{-+}=-1$) and the Higgs fields are represented as symmetric matrices. 

It is also often useful to represent $Z$ as a pair of $2 \times 2$ matrices $X$ and $Y$ and the Higgs fields as traceless matrices by raising some indices:
\begin{align}
X_\alpha^{\beta'} &= Z_{\alpha \beta +} \epsilon^{\beta \beta'} \cr
Y_\beta^{\alpha'} &= Z_{\alpha \beta -} \epsilon^{\alpha \alpha'} 
\end{align}
which allows us to write 
\begin{align}
\Phi &= X Y - \frac{1}{2} \mathrm{Tr} XY \cr
\Phi' &= -Y X + \frac{1}{2} \mathrm{Tr} XY \cr 
\Phi'' &= \begin{pmatrix}\frac{1}{2} \mathrm{Tr} XY & \det X \cr - \det Y & - \frac{1}{2} \mathrm{Tr} XY\end{pmatrix}
\end{align}

The first consequence of these equations is that the three Higgs fields have the same characteristic polynomial, 
i.e. the Lagrangian submanifold projects down to the tri-diagonal 
in the bases of the three Hitchin systems: 
\begin{equation}
\det \Phi = \det \Phi' = \det \Phi''
\end{equation}
In terms of the Hamiltonians $u_i^{(a)}$ of the integrable system, that means $u_i^{(a)}= u_i^{(b)}$ for all pairs $a$, $b$. 

Furthermore, we have intertwining relations such as 
\begin{equation}
\Phi X + X \Phi' = 0 \qquad \qquad Y \Phi + \Phi' Y =0
\end{equation}
and cyclic rotations thereof. 
As the Hamiltonian flows along the fibres of the integrable system can be understood as
shifts of $D_{\bar z}$ by monomials in the $\Phi^{(a)}$, the intertwining relationships show that 
the Lagrangian submanifold is invariant under differences of flows in pairs of copies,
generated by Hamiltonians $u_i^{(a)}- u_i^{(b)}$. In terms of angular coordinates $\theta^{(a)}$ on a generic  
fiber, the Lagrangian should be defined locally by equations of the form $\theta^{(1)}+\theta^{(2)}+\theta^{(3)}=f(u)$.

It is easy to see that the Lagrangian includes the tri-diagonal of the Hitchin section of the 
integrable system, associated to the same spin bundle $K^{1/2}$ used in the definition of $Z$: 
if the $SU(2)$ bundles are all $K^{1/2} \oplus K^{-1/2}$, then the $8$ components of $Z$
are sections of $K^{1/2 \pm 1/2 \pm 1/2 \pm 1/2}$. We can take as the only non-zero components
\begin{equation}
Z_{+++} = \phi(z) \qquad \qquad Z_{+--} = Z_{-+-} = Z_{--+} = 1
\end{equation}
i.e.
\begin{align}
X&= \begin{pmatrix}0 & -1 \cr \phi(z) & 0 \end{pmatrix} \cr
Y&=\begin{pmatrix}-1 & 0 \cr 0 & 1 \end{pmatrix} 
\end{align}
to get the canonical Hitchin section for each Higgs field:
\begin{equation}
\Phi = \Phi' = \Phi''= \begin{pmatrix}0 & 1 \cr \phi(z) & 0 \end{pmatrix} \end{equation}

Following the analysis of  \cite{Hitchin:2016aa} or proceeding analogously to our previous examples, 
one may conclude that at above a generic point in the bases of the Hitchin system, the complex Lagrangian is defined by the equations
\begin{equation}
u_i^{(1)}= u_i^{(2)}= u_i^{(3)} \qquad \qquad \theta^{(1)}+\theta^{(2)}+\theta^{(3)}=0
\end{equation}
where ``0'' is defined in terms of the Hitchin section for a given choice of $K^{1/2}$. 

\subsection{BBB analysis of tri-fundamental}
In order to complete the BBB analysis, we need to understand how to define the Fock space for 
a situation where the representation $M$ does not split. As the system should not have an anomaly, the 
problem should not be too hard. 

The Dirac-Higgs bundle for the tri-fundamental representation is generically a sum of lines 
defined by the cokernel of the map $\Phi \oplus \Phi' \oplus \Phi''$ on the tri-fundamental bundle. 
The map has rank $8$ away from points on $C$ where $P = P_1^2 + P_2^2 + P_3^2 - 2 P_1 P_2 - 2 P_1 P_3 - 2 P_2 P_3$ 
vanishes, where $P_i$ are the determinants of the Higgs fields. It is not obvious, but at these loci the rank drops by $2$. 
These are the loci where the Higgs field eigenvalues satisfy $\lambda \pm \lambda' \pm \lambda''=0$.

As the rank drops by $2$ at the special loci, each will contribute a two-dimensional sub-bundle. 
Thus the rank of the Fock space should be $2^n$, where $n=8g-8$ is the number of zeroes of $P$. 

In order to describe the sheaf of ground-states we need to compute the symplectic form on the 
Dirac-Higgs bundle for the tri-fundamental representation and seek a Lagrangian splitting. 
We will not attempt to do so, although we may observe that the images of 
holomorphic sections of $E \otimes E' \otimes E'' \otimes K$ modulo the image of $\Phi \oplus \Phi' \oplus \Phi''$
should give a Lagrangian subbundle of the Dirac-Higgs bundle. 

\subsection{BAA analysis of tri-transparent interface}
As the gauge fields for the three theories are identified, we must have 
\begin{equation}
E=E'=E''
\end{equation}
Dually, the Higgs fields must satisfy 
\begin{equation}
\Phi + \Phi' + \Phi'' = 0
\end{equation}

We can visualize the Higgs fields on each point of $C$ as three vectors in $\C^3$ forming a triangle. 
The square lengths $P_i$ of the vectors determine the geometry of the triangle, which collapses 
at the points on $C$ where  $P = P_1^2 + P_2^2 + P_3^2 - 2 P_1 P_2 - 2 P_1 P_3 - 2 P_2 P_3$ 
vanishes. 

It should be possible to show that there are $2^n$ possible choices of bundle $E$ given the $P_i$ and 
match them with the S-dual description. We leave this for future work. 

\subsection{BBB analysis of tri-transparent interface}
The interface should be the structure sheaf of the tri-diagonal
in the product of the three Hitchin systems. 
This is compatible with the BAA analysis of the dual interface. 

\section{More examples with gauge group reductions}
There is a large class of examples of simple interfaces whose S-dual description involves 
some kind of reduction of the gauge group to a subgroup. 
We review the properties of interfaces with gauge group reduction in Appendix \ref{sec:dir}.

\subsection{Abelianization interface between $SU(2)$ and $U(1)$ gauge theories}
Consider an interface between $SU(2)$ and $U(1)$ gauge theories, both with Neumann boundary conditions, coupled to 
a set of hypermultiplets in the fundamental representation of $SU(2)$, of charge $1$ under $U(1)$. 

This interface is expected to be S-dual to an interface where the $SU(2)$ gauge group is reduced to $U(1)$ 
and identified with the $U(1)$ gauge group on the other side. 

\subsubsection{BAA twist}
The equations
\begin{equation}
\Phi_{SU(2)} = X Y - \frac{1}{2} \, Y \cdot X\qquad \qquad \Phi_{U(1)} = Y \cdot X
\end{equation}
imply that $\mathrm{Tr} \Phi_{SU(2)}^2 = \frac{1}{4} \Phi_{U(1)}^2$. Hence the BAA brane is supported on the locus where the eigenvalues of the
$SU(2)$ Higgs field are globally defined. 

This agrees with the proposed mirror BBB brane, which is a sheaf supported on the graph of the diagonal embedding of $U(1)$ 
Hitchin systems into $SU(2)$ Hitchin systems, but makes the analysis more complicated and brings us outside the scope of a sigma model analysis.

Notice that the sections $X$ and $Y$ are globally defined left and right eigenvectors of $\Phi_{SU(2)}$ with eigenvalue $\frac{1}{2} \Phi_{U(1)}$. 
As discussed in previous examples, these relations imply that the Lagrangian is invariant under flows generated by the Hamiltonian
$\mathrm{Tr} \Phi_{SU(2)}^2 - \frac{1}{4} \Phi_{U(1)}^2$, as it should. 

\subsubsection{BBB twist}
The Dirac-Higgs bundle over the product of Hitchin moduli spaces, 
in the doublet representation of $SU(2)$, charge $1$ representation of $U(1)$,
can be described most easily at loci where $\Phi_{SU(2)} + \Phi_{U(1)}$ has generically full rank, 
dropping to rank $1$ at $4g-4$ isolated points in $C$ where $\mathrm{Tr} \Phi_{SU(2)}^2 = \frac{1}{4} \Phi_{U(1)}^2$.
Then the bundle is a sum of the co-kernels of $\Phi_{SU(2)} + \Phi_{U(1)}$.

The associated Fock bundle will thus have rank $2^{4g-4}$. This agrees with the expected mirror: 
the BAA twist of the S-dual interface is supported on the locus where $E_{SU(2)} = E^{\frac12}_{U(1)} \oplus E^{-\frac12}_{U(1)}$
and
\begin{equation}
\Phi_{SU(2)} = \begin{pmatrix} \frac{1}{2} \Phi_{U(1)} & U \cr -V & - \frac{1}{2} \Phi_{U(1)} \end{pmatrix}
\end{equation}
where $U$ and $V$ are sections of $K E_{U(1)}$ and $K E^{-1}_{U(1)}$. Thus we have 
\begin{equation}
\mathrm{Tr} \Phi_{SU(2)}^2 - \frac{1}{4} \Phi_{U(1)}^2 = U V
\end{equation}
and we can solve the problem by distributing the zeroes of $\mathrm{Tr} \Phi_{SU(2)}^2 - \frac{1}{4} \Phi_{U(1)}^2$ 
among $U$ and $V$, thus fixing $E_{U(1)}$. 

Thus the mirror BAA brane intersects the fiber of the complex integrable systems generically at $2^{4g-4}$
points. The $d$ and $q$ charges match as well, as for the $U(1)$ example. We leave as an exercise to confirm the fiberwise mirror identification of the 
points in the fiber with the line bundles of the original BBB brane.

\subsection{Half-fundamental for $SU(2)$}
As the fundamental representation of $SU(2)$ is pseudo-real, one can consider fundamental half-hypermultiplets.
These cannot be coupled to three-dimensional gauge fields due to a $Z_2$ anomaly, but can be coupled 
at a Neumann boundary condition or matter interface to four-dimensional gauge fields. The anomaly, though,
still has to be cancelled by anomaly inflow from a non-trivial discrete $\theta$ angle. 

In other words, the four-dimensional gauge theory compatible with coupling to a single fundamental half-hypermultiplet
at a Neumann boundary condition must be the theory indicated sometimes as $Sp(1)'$, which is mapped to itself 
under S-duality, rather than being mapped to $SO(3)$ gauge theory. 

Furthermore, the S-duality map between boundary conditions is also affected by this. The interface we are discussing here 
is S-dual to a maximal Nahm pole for the $Sp(1)'$ gauge theory. 

This setup generalizes to $Sp(n)'$ gauge theory for all $n$: Neumann boundary conditions with half-fundamental matter are dual to 
a maximal Nahm pole for $Sp(n)'$ gauge theory. 

The discrete theta angle must modify the Hitchin moduli space, by coupling the connection on $C$ to an appropriate gerbe. 
Roughly, this affects the way the Abelian fibers of the integrable system are glued together globally. 
The BAA and BBB branes must only make sense in the modified moduli space.

 An accurate analysis goes beyond the scope of this paper. In this section we mostly outline some general 
perspective on the problem. 

\subsubsection{BAA twist}
A mathematical discussion of this system can be found in \cite{Hitchin:2016aa}.

The BAA brane is supported on the origin of the base of the complex integrable system, $\mathrm{Tr} \Phi^2 =0$, 
as we have 
\begin{equation}
\Phi = X \, X\wedge
\end{equation}
This is compatible with the expected mirror, but brings us outside the scope of a sigma model analysis.  

\subsubsection{BBB twist}
The Dirac-Higgs bundle in the fundamental representation for general $\Phi$ receives contributions from the $4g-4$ points on $C$ 
where $\mathrm{Tr} \Phi^2$ vanishes. Generically, the rank of $\Phi$ only drops by $1$ at these loci. 
Thus the rank of the Dirac-Higgs bundle is generically $4g-4$. 

The associated Fock space bundle should have rank $2^{2g-2}$. As we are dealing with an half-hypermultiplet, 
in order to describe the Fock space bundle we need to figure out the symplectic form on the 
Dirac-Higgs bundle and pick a Lagrangian splitting. We will not attempt to do so here. 

The mirror BAA brane should be some version of the Hitchin section 
appropriate for the twisted Hitchin system. 

\subsection{Real forms}
There is a neat class of boundary conditions with gauge group reductions, 
which reduce the gauge group $G$ to a subgroup $H$ which is the maximally compact subgroup of 
some non-compact real form $G'$ of $G$. In particular, the BAA twist of such boundary conditions 
has a neat description in the complex structure where the holomorphic data is the complexified flat connection $d_A + \Phi + \bar \Phi$:
the connection lies in $G'$. The S-dual boundary conditions are known in many examples, see Table $3$ in \cite{Gaiotto:2008ak}.

In particular, if $G = SU(2N)$ and $H = S(U(N) \times U(N))$, 
so that the real form is $G' = SU(N,N)$, the S-dual boundary condition involves a gauge group reduction to 
$Sp(2N)$ coupled to a set of fundamental boundary hypermultiplets. 

Thus the BAA brane associated to $SU(N,N)$ should be mirror of the BBB brane 
supported on the $Sp(2N)$ Hitchin moduli space inside $SU(2N)$, given by the Fock space bundle 
built from the fundamental Dirac Higgs bundle for $Sp(2N)$. This is precisely the proposal of section 7 in 
\cite{2013arXiv1308.4603H}.

\section{D-modules}
\subsection{Hypermultiplets and symplectic bosons}
Following the finite-dimensional analogy in Appendix \ref{app:BAA}, the quantization of the Lagrangian submanifold $\CL(C,G,M)$ 
should be associated to the path integral 
\begin{equation}
\int DZ e^{\frac{1}{\hbar} \int_\Sigma \langle Z, \bar \partial_A Z \rangle}
\end{equation}
This is a well known object: it defines the partition function of a set of chiral symplectic bosons 
in two dimensions. 

In the absence of zeromodes, conformal blocks for these chiral symplectic bosons provides 
an interesting (twisted) rank 1 D-module, where the vector field $\hbar \frac{\delta}{\delta A_{\bar z}}$ 
acts as the insertion of a current 
\begin{equation}
J = : \mu(Z) :
\end{equation}

Concretely, the result of the path integral, which is a flat section for that D-module, is the inverse of the determinant 
of the $\bar \partial_A$ operator valued in $N$ (or square root of the determinant of the Dirac operator valued in $M$), which is 
a certain theta function $\Theta_M(A)$. In particular, the D-module has regular singularities on the locus where 
the $\bar \partial_A$ operator admits global sections, which is the locus where the classical 
description of the BAA brane has components with $\Phi \neq 0$.  

The space of conformal blocks in the neighbourhood of that singular locus is somewhat intricate. We refer the reader to \cite{Next} for a more in-depth discussion.

\subsection{Boundary gauge theories}
Following the finite-dimensional analogy in Appendix \ref{app:BAA}, the quantization of the Lagrangian submanifold $\CL(C,G,G_B,M)$ 
should be associated to the path integral 
\begin{equation}
\int DZ D\bar A_B e^{\frac{1}{\hbar} \int_\Sigma \langle Z, \bar \partial_{A+ A_B} Z \rangle}
\end{equation}
The meaning of this path integral is not immediately obvious, but a natural conjecture 
is that it should produce the conformal blocks of a coset 
\begin{equation}
\frac{\{ Z_M \}}{\hat G_B}
\end{equation}
of the symplectic boson theory by the $\hat G_B$ WZW current algebra 
generated by the moment maps $\mu_B(Z)$. 

We will pursue this idea at length in \cite{Next}. It passes some rather non-trivial tests. 

\section{Sheafs on the moduli space of local systems}
\subsection{Hypermultiplets}
The analysis of the Dirac-Higgs bundle in complex structure $J$ is rather straightforward: 
it gives us the sheaf $H^*_{\mathrm de Rahm}(C,G,M)$ over $\Loc(C,G)$ of de Rahm cohomology for forms on $C$ valued in 
$M$ (seen as a local system over $C$).

The cohomology is generically located at degree $1$ and the corresponding sheaf $\CV(C,G,M)$ over $\Loc(C,G)$ is given by the Fock space 
built from that degree 1 cohomology. If $M = N \oplus N^*$, we can write 
\begin{equation}
\CV(C,G,M) = (\det H^1_{\mathrm de Rahm}(C,G,N))^{-\frac12} \Lambda^* H^1_{\mathrm de Rahm}(C,G,N)
\end{equation}

\subsection{Boundary gauge theories}
We expect that in the presence of boundary gauge fields, the 
sheaf on $\Loc(C,G)$ will be the convolution of the sheaf $\CV(C,G \times G_B,M)$ on 
$\Loc(C,G) \times \Loc(C,G_B)$ built from the boundary matter fields and the structure sheaf on 
$\Loc(C,G_B)$.

Again, we stress that this picture may miss some important contributions
from the singular locus of $\Loc(C,G_B)$.

  \section*{Acknowledgements} 
We thank Kevin Costello and Edward Witten for many illuminating conversations and explanations. The research of DG was supported by the Perimeter Institute for Theoretical Physics. Research at Perimeter Institute is supported by the Government of Canada through Industry Canada and by the Province of Ontario through the Ministry of Economic Development \& Innovation.

\appendix

\section{Lagrangian submanifolds and generating functions}\label{app:BAA}
Consider the complex symplectic manifold $T^*B$ for some complex manifold $B$, with local coordinates $q$ along $B$ and $p$ along the fibre.
The complex symplectic form is $dp dx$ and the simplest complex Lagrangian submanifold $\CL_0$ is defined by $p=0$. 

Given an holomorphic superpotential $W(q)$ defined on $B$, we can define a complex Lagrangian submanifold $\CL_W$ as 
\begin{equation}
p = \frac{\partial W}{\partial q}
\end{equation}
This is Lagrangian because 
\begin{equation}
dp dq|_{\CL_W} = \frac{\partial^2 W}{\partial q^2} dq dq =0
\end{equation}

We can extend such construction by looking at a superpotential which depends on some auxiliary variables. We could take a superpotential 
function on $U \times B$, for some auxiliary complex manifold $U$ with coordinates $u$, or perhaps even defined on some complex fibration $F \to B$. 
We can consider the equations 
\begin{equation}
p = \frac{\partial W}{\partial q} \qquad \qquad \frac{\partial W}{\partial u}=0
\end{equation}
defining a submanifold $F_W$ in $F$. The projection of such submanifold onto $B$ is a complex Lagrangian $L_W$:
\begin{equation}
dp dq|_{L_W} = \frac{\partial^2 W}{\partial q^2} dq dq + \frac{\partial^2 W}{\partial q \partial u} dq du=- \frac{\partial^2 W}{\partial^2 u} du du =0
\end{equation}

The fibration $F_W$ can be used to produce certain locally constant sheaves over $L_W$. It is particularly natural to consider the cohomology
of the fiber.

This construction is compatible with symplectic quotient operations as long as $W$ is invariant under the group one quotients $T^*M$ by.
\subsection{Quantization}
There is a natural quantization of the Lagrangian submanifolds defined above. We refer to the reader to \cite{Next} for a more in-depth discussion.

In the case of an holomorphic superpotential $W(q)$ defined on $B$, we can define a D-module on $B$ whose sections are holomorphic functions on $B$, with 
\begin{equation}
\hat p f(q) = \hbar \partial_q f(q) + \frac{\partial W}{\partial q} f(q)
\end{equation}
This D-module encodes the behaviour of functions of the form $f(q) \exp \frac{W}{\hbar}$.

More generally, we can define a D-module on $B$ whose sections are holomorphic functions on $F$, modulo the subspace of functions of the form   
\begin{equation}
\hbar \partial_u g(q,u) + \frac{\partial W}{\partial u} g(q,u)
\end{equation}
with vector fields on $B$ acting by 
\begin{equation}
\hat p f(q,u) = \hbar \partial_q f(q,u) + \frac{\partial W}{\partial q} f(q,u)
\end{equation}

This D-module encodes the behaviour of functions of the form
\begin{equation}
\oint_U f(q,u) \exp \frac{W}{\hbar}
\end{equation}

\section{Supersymmetric Berry connections for $\CN=4$ SQM}\label{app:SQM}
\subsection{Supersymmetric Berry connection}
Consider a Supersymmetric Quantum Mechanics, defined by two supercharges $Q$ and $Q^\dagger$,
with $Q^2=0$ and Hamiltonian 
\begin{equation}
H = \{Q, Q^\dagger \}
\end{equation}
Supersymmetric ground states are annihilated by both super-charges. If the system has a gap, 
supersymmetric ground states have an equivalent description as the $Q$-cohomology of the Hilbert space. 

Next, consider a complex family of $\CN=2$ SQM systems, where the supercharge $Q(u)$ depends holomorphically 
on some parameters $u^a$ and $Q^\dagger(\bar u^a)$ depends anti-holomorphically on them. Then the ground states form 
a holomorphic sheaf on the parameter space: the anti-holomorphic derivatives $\partial_{\bar u^a}$ commute with 
the supercharge and each other and descend to commuting anti-holomorphic derivatives $D_{\bar u^a}$ on the cohomology
of $Q(u)$. We denote these parameters as ``B-type'' parameters, for reasons which will become apparent momentarily. 

Concretely, if we have some local basis $|i\rangle$ of the space of ground states, with $\langle i| |j \rangle = \delta_i^j$,
the Berry connection is defined by
\begin{equation}
(A_a)_i^j = \langle i|\partial_{u^a} |j \rangle \qquad (A_{\bar a})_i^j = \langle i|\partial_{\bar u^a} |j \rangle
\end{equation}
Thus along B-type directions, the curvature of the Berry connection is of type $(1,1)$. 

Another interesting situation is a real family of $\CN=2$ SQM systems, where the supercharge $Q(t)$ depends on
parameters $t^i$ in such a way that 
\begin{equation}
\partial_{t^i} Q(t) = [Q(t),T_i] \qquad \qquad \partial_{t^i} Q^\dagger(t) = -[Q^\dagger(t),T_i]
\end{equation}
for some Hermitean operators $T_i$, with $[\partial_{t^i}+T_i, \partial_{t^j}+T_j]$ being $Q(t)$-exact. 
Then the differential operators $\partial_t + T$ commute 
with the supercharge and descend to a complexified flat connection $D_t$ on the sheaf of ground states. 
We denote these parameters as ``A-type'' parameters. 

Thus along A-type directions, the Berry connection can be complexified to a flat supersymmetric Berry connection. 

More generally, we can have some family $Q(u,t)$ depending on both types of operators, which will lead to 
a sheaf of ground states admitting two commuting structures: a holomorphic Berry connection $D_{\bar u}$
and a flat complexified Berry connection $D_t$. 

\subsubsection{De Rahm SQM}
The simplest example of supersymmetric quantum mechanics 
takes the Hilbert space to consist of $L^2$ normalizable forms 
on a compact Riemannian manifold $X$ and the supercharge to be the exterior derivative:
\begin{equation}
Q = d \qquad Q^\dagger = d^\dagger
\end{equation}
so that the Hamiltonian is the Laplace operator on forms. The space of ground states coincides with the de Rahm cohomology of $X$. 

In this gauge the metric dependence is all in $Q^\dagger$ and if we have some parameter space of metrics on $X$,
the de Rahm cohomology is naturally a flat bundle over the parameter space. This is a rather trivial 
example of a real family of $\CN=2$ SQM systems, as the Berry connection is just flat, without need of complexification: $T$ vanishes. 

Morse-Witten quantum mechanics provides a somewhat more interesting example: 
\begin{equation}
Q = d + dh \wedge \qquad Q^\dagger = d^\dagger + (dh \wedge)^\dagger
\end{equation}
Here a variation of the Morse function leads to 
\begin{equation}
\partial_t Q = d(\partial_t h) \wedge = [Q,\partial_t h]
\end{equation}
and we get a complexified flat connection on spaces of Morse functions. 

If we were to replace $dh$ with a more general closed 1-form, the variation of $Q$ along the parameter space 
would not be exact. We can get a holomorphic Berry connection if we enlarge the parameter space by including also 
a twist by a $U(1)$ connection on $X$: 
\begin{equation}
Q = d_A \qquad Q^\dagger = d_A^\dagger 
\end{equation}
for some complex (i.e. $GL(1)$) flat connection $A$. The anti-Hermitean part of $A$ is the $U(1)$ connection,
while the Hermitean part of $A$ is the generalization of the ``$dh\wedge$'' part of the Morse quantum mechanics. 

Notice that unitary gauge transformations leave the quantum mechanics unchanged, while complexified gauge transformations 
are equivalent to adding a Morse function to the system. The parameter space will be a space of 
$GL(1)$ flat connections on $X$ modulo unitary gauge transformations, to be thought of as a fibration over a parameter space 
of $GL(1)$ flat connections on $X$ modulo $GL(1)$ gauge transformations, with fibres being spaces of Morse functions. 

The space of $GL(1)$ flat connections on $X$ modulo $GL(1)$ gauge transformations is naturally a complex manifold 
and $Q$ varies holomorphically on it. Thus we obtain a supersymmetric Berry connection 
which is holomorphic along the space of $GL(1)$ flat connections on $X$ modulo $GL(1)$ gauge transformations
and flat along the fiber directions of the space of $GL(1)$ flat connections on $X$ modulo unitary gauge transformations.

This system has obvious generalizations where $X$ is equipped with vector bundle and a complexified flat connection.

More generally, one can consider some auxiliary quantum mechanics with $X$ as a parameter space of A-type deformations and write the combined supercharge
\begin{equation}
 \epsilon (d_x + T \wedge ) + Q(x)
\end{equation}
The A-type constraints on $T$ and $\partial_x Q(x)$ guarantee that the above supercharge is nilpotent. In the Born-Oppenheimer approximation 
for small $\epsilon$, the ground states of the system can be computed by the reduced super-charge $d_A$ where $A$ is the supersymmetric Berry connection 
for $Q(x)$.

\subsubsection{Dolbeault SQM}
Another natural example of supersymmetric quantum mechanics involves a complex manifold $Y$ equipped with a holomorphic vector bundle $E$,
both endowed with an Hermitean metric. 
We can define 
\begin{equation}
Q = \bar \partial_E \qquad Q^\dagger =(\bar \partial_E)^\dagger
\end{equation}
acting on $(0,q)$ forms valued in $E$. The $Q$ cohomology is the Dolbeault cohomology of $E$. 

The supercharge and cohomology depend holomorphically on the choice of vector bundle $E$ and trivially on the choice 
of Hermitean metrics. 

There is an useful extension to a complex of vector bundles $E^\cdot$, with some holomorphic differential $\delta$
and supercharge $Q = \bar \partial_{E^\cdot} + \delta$. The space of ground states is the {\it hypercohomology} of the complex. 

It is also interesting to consider some auxiliary quantum mechanics with $Y$ as a parameter space of B-type deformations and write the combined supercharge
\begin{equation}
 \epsilon \bar \partial_y + Q(y)
\end{equation}
In the Born-Oppenheimer approximation for small $\epsilon$, the ground states of the system can be computed by the reduced super-charge $\bar \partial_E$ where $E$ is the sheaf of ground states for $Q(y)$.
\subsubsection{From SQM to branes}
The $\CN=2$ SQM systems discussed above can be used as boundary degrees of freedom in defining 
branes for two-dimensional sigma models. In particular, if the gap of the quantum mechanical system is 
much larger than the scale at which the sigma model becomes strongly coupled, the space of ground states 
of the quantum mechanical system becomes the Chan-Paton bundle for a traditional UV boundary condition for the 
sigma model. \footnote{It is also reasonable to consider situations where a finite number of non-ground states 
survive at the sigma model UV scale, giving rise to Chan-Paton complex equipped with a differential. }

The natural way to define such coupling to a sigma model is to start from some reference boundary condition for the sigma model and
then promote the boundary values of the bulk fields to parameters in the SQM. 

B-type boundary conditions involve the promotion of B-type deformation parameters in the SQM. For example, we can start from 
some reference Neumann boundary condition associated to the structure sheaf of the sigma model target space $U$ 
and couple it to a SQM with $U$ as a parameter space of B-type deformations. 

The result should be a standard B-brane based on the holomorphic sheaf of ground states of the SQM over $U$.  

A-type boundary conditions involve the promotion of A-type deformation parameters in the SQM. 
It is instructive to first consider the example of a Morse-Witten SQM. 

Consider a sigma model on $T^*X$, with the standard boundary condition given by the base $X$ of the co-tangent bundle. In local coordinates $(x,p)$ 
on $T^*X$ we can denote it as $p=0$.
 
Consider a Morse-Witten SQM on some auxiliary manifold $Y$ with a family of Morse functions $h(x,y)$ on $Y$ parameterized by $X$. 
Classically, when $h$ has isolated critical points as a function of $y$, 
the boundary coupling induced by $h(x,y)$ deforms the boundary condition to $p=\partial_x h(x,y^*)$, 
where the Morse function is evaluated at a critical point $y^*$ solving $\partial_y(x,y^*)=0$. This is a 
Lagrangian submanifold which is a multi-cover of $X$ in $T^*X$. It supports a flat $U(1)$ bundle 
given by the cohomology class associated to the critical point.

A better way to think about this is an A-brane which is a deformation of multiple copies of $X$ 
by the complexified Berry connection $D_x$: the unitary part of the connection 
is the connection of the CP bundle on $X$, while the Hermitean part $T$ 
should be understood as an Higgs field, a matrix-valued version of a deformation of the Lagrangian manifold 
along the normal direction to $X$. When $T$ and $A$ are approximately diagonal this 
reduces to the multi-cover of $X$ mentioned above. 

This perspective applies in general to any SQM with $X$ as a space of A-type deformations. 

\subsubsection{From branes so SQM}
Conversely, a nice example of families of $\CN=2$ SQM arise when 
one considers the compactification of a $(2,2)$ sigma model on a segment, 
with families of supersymmetric boundary conditions at the endpoint. 

As an example, consider families of A-branes. An A-brane described by a Lagrangian 
submanifold equipped with a flat connection has a moduli space of deformations 
given by deformations of the manifold or deformations of the flat connection. 

A deformation of the Lagrangian submanifold is a section of the normal bundle, 
which can be mapped to a closed 1-form with the help of the symplectic form. 

If the 1-form is exact, we obtain an Hamiltonian function which generates the 
deformation of the Lagrangian manifold. If the 1-form is not exact, 
then it gives us a flat bundle deformation. 

Thus general deformations are complex, combining corresponding shape and flat connection deformations.
They are actually B-type deformations of the SQM. On the other hand, exact shape deformations 
are A-type deformations, with the Hamiltonian playing the role of $T$. 

\subsection{Hyper-holomorphic Berry connections}
Next we can consider quantum-mechanical systems with ${\cal N}=4$ supersymmetry.
In the simplest situation, we have four Hermitean supercharges $\tilde Q_i$ which satisfy 
\begin{equation}
\{\tilde Q_i, \tilde Q_j\} = \delta_{ij} H
\end{equation}
The supersymmetric ground states of the system are annihilated by all supercharges.

It is also useful to consider complex supercharges $Q_1 = \tilde Q_1 + i \tilde Q_2$ and $Q_2 = \tilde Q_3 + i \tilde Q_4$, 
so that in particular 
\begin{equation}
\{Q_\alpha, Q_\beta\} =0 
\end{equation}
and furthermore 
\begin{equation}
\{Q_1, (Q_2)^\dagger \} =0 \qquad \{Q_1, (Q_1)^\dagger \}= \{Q_2,(Q_2)^\dagger \} =H
\end{equation}
Lowering the index of the second supercharge with an anti-symmetric tensor, we define $\bar Q_2 = (Q_1)^\dagger$ and $\bar Q_1 = - (Q_2)^\dagger$, so that 
\begin{equation}
\{Q_\alpha, \bar Q_\beta\} =\epsilon_{\alpha \beta} H 
\end{equation}
In particular, the supersymmetric ground states can be identified with the cohomology of $Q_1$, or the cohomology of $Q_2$, 
or the cohomology of any other linear combination $Q_\zeta = Q_1 + \zeta Q_2 = Q_\alpha \zeta^\alpha$. 

In a more ``twistorial'' language, the SUSy relations can be expressed in terms of $Q_\zeta$ and $\bar Q_\zeta = \bar Q_1 + \zeta \bar Q_2$ as 
\begin{equation}
Q_\zeta^2 = 0 \qquad \{Q_\zeta, \bar Q_\zeta \}=0 \qquad \bar Q_\zeta^2 = 0
\end{equation}
with a reality condition 
\begin{equation}
(\bar Q_\zeta)^\dagger = \zeta^\dagger Q_{- 1/\zeta^\dagger}
\end{equation}

We will describe various ways the Berry connection for these ground states can be affected by the $\CN=4$ supersymmetry. 
We will be mostly interested in two possibilities. BBB-type deformations are analogous to Nahm transforms and 
will give us a Berry connection which is hyper-holomorphic. BAA deformations are analogous to the $tt^*$ geometry of 
$(2,2)$ SQFTs \cite{Cecotti:1991me} and will give us a Berry connection which combined to a Higgs field is a solution of generalized Hitchin's equations.
There are other possibilities, which we will not need here, which give solutions of generalized Nahm or BPS equations \cite{Cecotti:2013mba}. 

A BAA-type deformation is in particular a B-type deformation for $Q_1$ and for $\bar Q_1 = Q_2^\dagger$ which is also an A-type deformation for all other $Q_\zeta$. 
That means we have $\partial_{\bar u} Q_1 = 0$ and $\partial_u Q_2=0$, but 
\begin{equation}
\partial_u Q_1 = \partial_u (Q_1 + \zeta Q_2) = [Q_1 + \zeta Q_2, T_u(\zeta)] \qquad \qquad \zeta \partial_{\bar u} Q_2 = [Q_1 + \zeta Q_2, T_{\bar u}(\zeta)]
\end{equation}
which we solve by 
\begin{equation}
\partial_u Q_1 = [Q_2, C_u] \qquad  [Q_1, C_u] =0 \qquad  \partial_{\bar u} Q_2 = [Q_1, C_{\bar u}(\zeta)]  \qquad  [Q_2, C_{\bar u}] =0
\end{equation}
We also require $[C_u, C_{\bar u}]=0$. 

These relations mean that the Lax operators
\begin{equation}
\partial_u + \frac{C_u}{\zeta} \qquad \qquad \partial_{\bar u} + \zeta C_{\bar u}
\end{equation}
commute and descend to a supersymmetric Berry connection which is flat for all values of $\zeta$, 
i.e. the Lax connection for a solution of Hitchin equations, with the Higgs field $\Phi$ being the projection of $C$ on ground states. 

For multiple deformation directions, we also require extra consistency conditions such as 
$[C_u, C_{u'}]=0$, $\partial_u C_{u'} = \partial_{u'} C_u$, $[C_u, C_{\bar u'}]=0$ in order to get a solution of 
generalized Hitchin equations on the parameter space. 

A BBB deformation is in particular a B-type deformation for all $Q_\zeta$, but in a different way for each $\zeta$. 
That means the parameter space $M$ is hyper-k\"ahler and $Q_\zeta$ is holomorphic in complex structure $\zeta$ 
on $M$. If we model locally the deformation space on ${\mathbb C}^{2n}$, 
so that the anti-holomorphic derivatives take the form 
\begin{align}
\partial_{\bar u} &+ \zeta \partial_v \cr
\partial_{\bar v} &- \zeta \partial_u 
\end{align} 
we get relations of the form 
\begin{equation}
\partial_{\bar u} Q_1 =0 \qquad \partial_v Q_1 + \partial_{\bar u} Q_2=0 \qquad \partial_v Q_2=0
\end{equation}
and
\begin{equation}
\partial_{\bar v} Q_1 =0 \qquad \partial_u Q_1 - \partial_{\bar v} Q_2=0 \qquad \partial_u Q_2=0
\end{equation}
More generally, if we decompose the tangent bundle of $M$ as the tensor product of 
an $SU(2)$ and an $Sp(n)$ bundles, we would write 
\begin{equation}
\partial_{I (\alpha} Q_{\beta)} = 0
\end{equation}
The Berry connection has curvature of type $(1,1)$ in all complex structures and thus the ground states form a hyper-holomorphic sheaf on $M$. 

It is possible to have systems which admit both BAA and BBB deformations. Then the Lax connection along the BAA direction 
commutes with the anti-holomorphic derivatives in complex structure $\zeta$ along $M$.

Although we will not need it for the examples in this paper, the analysis can be generalized to 
situations with central charges. That means a more general supersymmetry algebra
\begin{equation}
\{Q_\alpha, Q_\beta\} =0  \qquad \qquad \{Q_\alpha, \bar Q_\beta\} =\epsilon_{\alpha \beta} H + Z_{\alpha \beta}
\end{equation}
where the triplet of central charges $Z_{\alpha \beta}$ commute with all other operators. 

In sectors with non-zero $Z_{\alpha \beta}$ we can only find BPS ground states with $E = |Z|$, annihilated by half of the supercharges only.  
In particular, they are annihilated by $Q_\zeta$ only if $\zeta^\alpha$ is an eigenvector of $Z_\alpha^\beta$ with eigenvalue $|Z|$. 
In the sector annihilated by the central charges $Z_{\alpha \beta}$, the supersymmetric ground states are annihilated by all 
supercharges and can be identified with the cohomology of any $Q_\zeta$. These are the states to which our discussion would apply. 

\subsection{Dirac operators}
Several constructions such as ADHM constructions and Nahm transforms, which are usually stated in terms of zeromodes of 
Dirac operators, arise naturally in the context of $\CN=4$ SQM. 

Typically, one has some Dirac operator 
\begin{equation}
\Delta = {A_1 \choose A_2}
\end{equation}
a linear map or differential operator between two spaces $V$ and $W \oplus W$, $V$ and $W$ being equipped with an Hermitean inner product, 
with the property that 
\begin{equation}
\Delta \Delta^\dagger = {A_1 \choose A_2} (A_1^\dagger \quad A_2^\dagger) = \begin{pmatrix}A_1 A_1^\dagger & A_1 A_2^\dagger \cr A_2 A_1^\dagger & A_2 A_2^\dagger \end{pmatrix} = \begin{pmatrix}h & 0 \cr0 & h \end{pmatrix}
\end{equation}
i.e. $A_\alpha \bar A_\beta =h \epsilon_{\alpha \beta}$. Here we lower indices with $\epsilon_{\alpha \beta}$ as before, so that $\bar A_2 = (A_1)^\dagger$ and $\bar A_1 = - (A_2)^\dagger$. 
Notice that $(\bar A_1)^\dagger = -A_2$ and $-(\bar A_2)^\dagger = -A_1$.

The zeromodes of $\Delta$ are then used produce the desired output, such as a hyper-holomorphic connection on parameter space, or a solution of Hitchin, Nahm or BPS equations. 
Concretely, zeromodes are elements of $V$ annihilated by both $A_\alpha$. In typical situations, there is some index theorem relating the number of zeromodes of $D$ and $D^\dagger$.
Furthermore, $D^\dagger$ has generically no zeromodes.  

We can assemble four supercharges from the linear maps in the Dirac operator, acting on the Hilbert space $W \oplus V \oplus W$:
\begin{align}
Q_\alpha: W \xrightarrow{-\bar A_\alpha} V \xrightarrow{A_\alpha} W \cr
\bar Q_\alpha: W \xleftarrow{A_\alpha} V \xleftarrow{\bar A_\alpha} W
\end{align}
Then $\{Q_\alpha, Q_\beta\} =0$ and 
\begin{equation}
\{Q_\alpha, \bar Q_\beta\} = \begin{pmatrix} h &0 &0 \cr 0 & A_1^\dagger A_1 + A_2^\dagger A_2 &0 \cr 0 & 0 & h   \end{pmatrix} \epsilon_{\alpha \beta}
\end{equation}

The zeromodes of $D$ coincide with the supersymmetric groundstates of the $\CN=4$ SQM which lie in $V$. 
On the other hand, supersymmetric ground states in either of the two $W$ summands would in particular be 
zeromodes of $D^\dagger$ and will generically be absent.

The zeromodes or ground states can then be identified with the cohomology of $Q_\zeta$ for any $\zeta$. 
The dimension of the cohomology is generically $\dim V - 2 \dim W$. 

\subsubsection{Quantization of Dirac zeromodes}\label{sec:ham}
There are several physical situations, including one of the two main classes employed in this paper, 
where the complexes described in the previous section arise as a semi-classical phase space of a system. The actual 
$\CN=4$ quantum mechanical system involves a Fock space built from these complexes. 

Consider a quantum mechanical system involving bosonic creation operators $(x_+, y_+)$ valued in $W \oplus W$ and 
fermionic creation operators $\xi_+$ valued in $V$. We denote the corresponding destruction operators as $(x_-, \xi_-, y_-)$. 
We can thus define supercharges 
\begin{align}
Q_\alpha: -\xi_+ \bar A_\alpha x_- + y_+ A_\alpha \xi_- \cr
\bar Q_\alpha: x_+ A_\alpha \xi_- +  \xi_+ \bar A_\alpha y_-
\end{align}
so that the Hamiltonian becomes $H =  x_+ h x_- + y_+ h y_- + \xi_+  (A_1^\dagger A_1 + A_2^\dagger A_2) \xi_-$. 

This is a nice, free quantum mechanical system whose supersymmetric ground states are essentially the exterior algebra of 
the space of zeromodes in the previous section. More precisely, if we denote the space of zeromodes in $V$ as $Z$, 
the supersymmetric ground states of the quantum mechanics become the fermionic Fock space
\begin{equation}
FZ \equiv (\det Z)^{-\frac12} \Lambda^* Z
\end{equation}

There is another important $\CN=4$ quantum mechanical system one may consider, which is the dimensional reduction of a $(0,4)$ 2d theory. 
It involves a set of $(0,4)$ hypermultiplets valued in $W \oplus W$ and $(0,4)$ Fermi multiplets valued in $V$. 
These can be coupled supersymmetrically to background $(0,4)$ twisted hypermultiplets $A_\alpha$. 
We can write some supercharges:
\begin{align}
Q_\alpha &= \bar p_y \eta_\alpha +  \bar \eta_\alpha \bar p_x + \psi \bar A_\alpha x + y A_\alpha \bar \psi \cr 
\bar Q_\alpha &=  p_y \bar \eta_\alpha - \eta_\alpha p_x -  \bar x A_\alpha \bar \psi +  \psi  \bar A_\alpha \bar y
\end{align}
with $\{\eta_\alpha, \bar \eta_\beta\} = \epsilon_{\alpha \beta}$ are fermions in $W$ and $\{\psi, \bar \psi\} = 1$ are fermions in $V$, $x$ and $y$ 
are complex bosons. The Hamiltonian is roughly
\begin{equation} 
|p_y|^2 + |p_x|^2 + h |x|^2 + h |y|^2 + \psi  \bar A_\alpha \eta^\alpha +  \bar \eta^\alpha A_\alpha \bar \psi 
\end{equation}
The gap is now determined by $\sqrt{h}$. As long as $h$ is positive definite, the ground states are again the Fock space $FZ$ 
built from the fermion zeromodes $\bar \psi$ annihilated by $A_\alpha$. Intuitively, the mass terms pair up the $\eta_\alpha$ 
and some $\psi$ modes and lift them. 

\subsubsection{The ADHM system}
The ADHM constructions of $U(N)$ instantons in $R^4$ will provide our first example of BBB-type deformations, as 
hyper-holomorphic bundles in real dimension $4$ are precisely instantons. 

The ADHM data consists of $k \times k$ matrices $B_1$ and $B_2$ and $k \times N$ and $N \times k$ matrices $I$ and $J$, 
satisfying appropriate moment map constraints for a $U(k)$ hyper-K\"ahler quotient. 

The ADHM construction employs as a Dirac operator the $2k \times (N + 2k)$ matrix 
\begin{equation}
\Delta = \begin{pmatrix}I & B_1 + z_1 & B_2 + z_2 \cr  J^\dagger & B^\dagger_2 + z^\dagger_2 & -B^\dagger_1 - z^\dagger_1 \end{pmatrix} 
\end{equation}
with $z_1$ and $z_2$ being holomorphic coordinates on $R^4$. 

The supercharge $Q_1$ thus takes the form 
\begin{equation}
Q_1: \C^k \xrightarrow{ \begin{pmatrix} J \cr B_2 + z_2 \cr -B_1 - z_1 \end{pmatrix}} \C^{2k+N} \xrightarrow{\begin{pmatrix}I & B_1 + z_1 & B_2 + z_2\end{pmatrix}} \C^k
\end{equation}
and the remaining supercharges are defined as above. The B-type parameter space of the SQM consists both of 
$R^4$ and of the moduli space $\CM_{\mathrm{ADHM}}$ of the ADHM data. Notice how the $U(k)$ hyper-K\"ahler quotient of the ADHM data 
happens here: the moment map constraints are required for the $\CN=4$ supersymmetry algebra to hold
and the $U(k)$ action on the ADHM data gives equivalent quantum mechanics.  

The supersymmetric ground states of the SQM give a hyper-holomorphic bundle on $R^4$ for every choice of ADHM data which is the instanton bundle itself. 
More generally, they give an hyper-holomorphic bundle on $R^4 \times \CM_{\mathrm{ADHM}}$ which is the universal bundle on the instanton moduli space. 

In any given complex structure, the corresponding holomorphic bundle is computed as the cohomology of the $Q_\zeta$ complex. 

\subsubsection{K\"ahler SQM}
The standard $\CN=2$ SQM sigma model has enhanced supersymmetry when the target space $X$ is K\"ahler:
the exterior derivative splits as $d = \partial + \bar \partial$ and we can use $\partial$, $\bar \partial$, $\partial^\dagger$ and $\bar \partial^\dagger$ 
as four supercharges for an $\CN=4$ SQM. 

If we set $Q_1 = \bar \partial$, we have two possible choices: we can take either $Q_2 = \partial$ or $Q_2 = \partial^\dagger$.
Although the two choices give the same quantum mechanics, the two choices allow one to discuss two distinct and incompatible classes of 
supersymmetric deformations. In particular, the first choice includes complex structure and superpotential deformations while the second 
choice includes deformations of the K\"ahler metric and equivariant deformations. The first choice is the most natural to describe BBB and BAA spaces of deformations. 
The second would give Nahm- and BPS-type deformations. 
 
At first, we can consider a moduli space $U$ of complex structure deformations of a compact K\"ahler target manifold $X$. 
The super-charge at $\zeta=1$ is simply the de Rahm differential. Thus the cohomology at $\zeta=1$ is simply the 
de Rahm  cohomology, which is independent of the complex structure and obviously gives a flat bundle on the space 
$U$ of complex structure deformations. On the other hand, at $\zeta=0$ we have the Dolbeault cohomology of $X$, which is a natural holomorphic bundle on $U$. 

To explore the system in further detail, consider the variation of the $\bar \partial$ operator under a change in complex structure. Obviously, $\delta_{\bar u} \bar \partial=0$. 
We can also write
\begin{equation}
\delta_u \bar \partial = [\partial,i_{\mu_u}]
\end{equation}
where $i_{\mu_u}$ is an operation of contraction with the Beltrami differential corresponding to the variation of complex structure along $u$. 
The Beltrami differentials also satisfy the expected $[\bar \partial, i_{\mu_u}]=0$ and the other relations we expect on $C_u$ and $C_{\bar u}$,
which follow from the definition as variations of the complex structure $J$. 

As a result, we learn that variations of complex structure of $X$ are BAA-type deformations and the Berry connection $(D_u, D_{\bar u})$ and the expectation values $\Phi_u$ and $\Phi_{\bar u}$ of 
the Beltrami differential operators $i_{\mu_u}$ and $i_{\mu_{\bar u}}$ solve the generalized Hitchin equations on $U$. In complex structure $\zeta=0$, the 
data reduces to the Higgs bundle given by the holomorphic bundle of Dolbeault cohomology of $X$ together with the operators of contraction with holomorphic Beltrami differentials. 
In complex structure $\zeta=1$ the data reduces to the bundle of de Rahm comonology equipped with the flat Gauss-Manin connection. 
In general complex structure, we have the cohomology of the $\bar \partial + \zeta \partial$ operator and associated complexified flat connection. 

Next, we can consider a non-compact K\"ahler target manifold $X$ equipped with a holomorphic super-potential $W$. 
The relevant supercharges are 
\begin{equation}
Q_1 = \bar \partial + \partial W \wedge \qquad \qquad Q_2 = \partial + \bar \partial \bar W \wedge 
\end{equation}
The moduli space of BAA deformations now include both complex structure deformations and changes in the super-potential.
The corresponding $C_u$ operators are simply multiplication by the holomorphic functions $\partial_u W$. 

If we replace $\partial W$ with a more general holomorphic $(1,0)$ form $\lambda$, only exact changes in  $\lambda$ will be associated to BAA deformations. 
General deformations can be promoted to BBB-type deformations by adding the extra data of a flat $U(1)$ connection $a$: 
\begin{equation}
Q_1 = \bar \partial_a + \lambda \wedge \qquad \qquad Q_2 = \partial_a + \bar \lambda \wedge 
\end{equation}
so that the variation of $Q_1$ under a change in $\lambda$ matches the variation of $Q_2$ under a change in $a$. 

More generally, given any solution $(A,\Phi, \bar \Phi)$ of generalized Hitchin's equations on $X$ we can define a corresponding quantum mechanics:
\begin{equation}
Q_1 = \bar \partial_A + \Phi \wedge \qquad \qquad Q_2 = \partial_A + \bar \Phi \wedge 
\end{equation}
The deformations of $(A,\Phi, \bar \Phi)$ are BBB-type deformations.  

In complex structure $\zeta=0$, the ground states are identified with the cohomology of $\bar \partial_A + \Phi \wedge$. In terms of the corresponding Higgs bundle $(E,\Phi)$, this is the 
 {\it hypercohomology} of a complex of vector bundles on $X$: the bundles of forms valued in $E$ with differential $\Phi \wedge$. 

If $X$ is a Riemann surface $C$ and $(A,\Phi, \bar \Phi)$ a solution of the standard Hitchin equations on $C$, the quantum mechanics above is associated to a Dirac operator 
\begin{equation} \label{eq:DH}
\Delta = \begin{pmatrix} \bar D_A & \Phi \cr \bar \Phi & -D_A \end{pmatrix}
\end{equation}
The space of supersymmetric ground states is thus identified with the space of zeromodes of $\Delta$, which in turn is known as the 
{\it Dirac-Higgs bundle}, a natural hyper-holomorphic bundle on the Hitchin moduli space. 

This general type of quantum mechanics arises universally as the Born approximation of a system of the form 
\begin{equation}
Q_1 = \bar \partial_x + C \wedge + Q^{\mathrm{aux}}_1(x) \qquad \qquad Q_2 = \partial_x + \bar C \wedge + Q^{\mathrm{aux}}_2(x)
\end{equation}
where the ``slow'' degrees of freedom $x \in X$ act as BAA-type deformation parameters of the auxiliary degrees of freedom 
described by $Q^{\mathrm{aux}}_\alpha(x)$. The constraint of BAA-type guarantee that the above supercharges define an $\CN=4$ quantum-mechanical system. 

\subsubsection{Hyper-holomorphic SQM}
Our second example of quantum-mechanical system is based on a hyper-K\"ahler target space $Y$ equipped with an hyper-holomorphic bundle $E$. 
It is precisely the same as the Dolbeault SQM, but the extra symmetries guarantee the enhancement of supersymmetry to $\CN=4$. 
In particular, besides $Q_1 = \bar \partial_E$ one finds a second $Q_2 = \tilde \partial^\dagger_E$ with the property that 
$Q_\zeta$ is equivalent to the Dolbeault differential $\bar \partial_E^\zeta$ in complex structure $\zeta$.
This relies on the identification between spaces of $(0,q)$ forms in different complex structure 
induced by the reduction of the structure group of the cotangent bundle to $SU(2) \times Sp(n)$ 
on hyperk\"ahler manifolds. 

For example, given an instanton bundle on $R^4$, we can consider the Dirac operator 
\begin{equation}
\Delta = \begin{pmatrix} -\bar D_1 & \bar D_2 \cr D_2 & D_1 \end{pmatrix}
\end{equation}
acting on $L^2$ normalizable functions on $R^4$.

The corresponding supercharge 
\begin{equation}
Q_1 : L^2(R^4) \xrightarrow{{\bar D_2 \choose \bar D_1}} L^2(R^4) \oplus L^2(R^4) \xrightarrow{(-\bar D_1 \quad \bar D_2)} L^2(R^4)
\end{equation}
can be identified with the $\bar \partial_E$ operator acting on $(0,q)$ forms. 
The general supercharge 
\begin{equation}
Q_\zeta : L^2(R^4) \xrightarrow{{\bar D_2 + \zeta D_1 \choose \bar D_1- \zeta D_2}} L^2(R^4) \oplus L^2(R^4) \xrightarrow{(-\bar D_1+ \zeta D_2 \quad \bar D_2+ \zeta D_1)} L^2(R^4)
\end{equation}
can be identified with the $\bar \partial^\zeta_E$ operator acting on $(0,q)$ forms in complex structure $\zeta$. Notice that the same middle term in the complex is identified with 
$(0,1)$ forms in different complex structures. 

Hyper-holomorphic bundles can admit all sort of different deformation moduli, including BBB-, BAA-, BPS- and Nahm- type deformation parameters. 
Indeed, canonical examples are universal bundles for moduli spaces of solutions of instanton, Hitchin, BPS and Nahm equations. 

This general type of quantum mechanics arises universally as the Born approximation of a system of the form 
\begin{equation}
Q_1 = \bar \partial_y + Q^{\mathrm{aux}}_1(y) \qquad \qquad Q_2 = \tilde \partial_y + Q^{\mathrm{aux}}_2(y)
\end{equation}
where the ``slow'' degrees of freedom $y \in Y$ act as BBB-type deformation parameters of the auxiliary degrees of freedom 
described by $Q^{\mathrm{aux}}_\alpha(y)$. The constraint of BBB-type guarantee that the above supercharges define an $\CN=4$ quantum-mechanical system. 

\subsubsection{Hypermultiplets on a Riemann surface}
The BBB twist of three-dimensional hypermultiplets valued in $M$ on a Riemann surface $C$ can be thought of as 
the dimensional reduction of four-dimensional hypermultiplets twisted on $C$, which is a system with 2d $(0,4)$ 
supersymmetry. 

The hypermultiplet scalars reduced on $C$ together with half of the fermions give $(0,4)$ hypermultiplets, 
while the other half of the fermions give $(0,4)$ Fermi multiplets. In the notation used in \ref{sec:ham},
$W$ is the space of $M$-valued functions on $C$ and $V$ is the space of real $M$-valued 1-forms on $C$,
where the reality conditions uses the symplectic pairing on $M$ and the symplectic pairing on 1-forms.

The background twisted hypermultiplets $A_\alpha$ are identified with the differential operators $\bar \partial_A + \Phi\wedge$ 
and $\partial_A + \bar \Phi\wedge$, so that the Dirac operator $\Delta$ coincides with the Dirac-Higgs operator
\ref{eq:DH}. The Hamiltonian for the $(0,4)$ SQM can be readily matched to the Hamiltonian for the 
hypermultiplets. 

As long as the $h$ built from the connection and Higgs field is positive definite, the ground states are the Fock space 
built from the fermion zeromodes annihilated by $A_\alpha$, i.e. the fermionic Fock space modelled on the Dirac-Higgs bundle.

If $h$ has zeromodes, then the system has bosonic flat directions and a continuum of states touching the ground states. 

\subsubsection{From $\CN=4$ SQM to branes}
The $\CN=4$ SQM systems discussed above can be used as boundary degrees of freedom in defining 
branes for two-dimensional sigma models with eight supercharges, generalizing the statement for $\CN=2$ SQM systems.

Again, the natural way to define such coupling to a sigma model is to start from some reference boundary condition for the sigma model and
then promote the boundary values of the bulk fields to parameters in the SQM. 

BBB-type boundary conditions involve the promotion of BBB-type deformation parameters in the SQM. For example, we can start from 
some reference Neumann BBB boundary condition associated to the structure sheaf of the sigma model target space $U$ 
and couple it to a $\CN=4$ SQM with $U$ as a parameter space of BBB-type deformations. 

The result should be a standard BBB-brane based on the hyper-holomorphic sheaf of ground states of the SQM over $U$.  

BAA-type boundary conditions involve the promotion of BAA-type deformation parameters valued in a complex manifold $X$, 
with $T^*X$ being endowed with some local hyperk\"ahler metric. 
We can consider a BAA-brane which is a deformation of multiple copies of the BAA brane $X$ in $T^*X$
by the Berry connection and Higgs field produced by the SQM: the connection 
is the connection of the CP bundle on $X$, while the Higgs field $C$ 
should be understood as a matrix-valued version of a complex deformation of the Lagrangian manifold 
along the normal direction to $X$. When the Higgs field and connection are approximately diagonal this 
reduces to a complex Lagrangian multi-cover of $X$. 

\subsubsection{From branes so SQM}
Conversely, a nice example of families of $\CN=4$ SQM arise when 
one considers the compactification of a $(4,4)$ sigma model on a segment, 
with families of supersymmetric boundary conditions at the endpoint. 

As a neat concrete example, consider families of BAA-branes in the cotangent bundle $T^*C$ to a Riemann surface $C$ \cite{witten}
An obvious family of BAA branes consist of the fibers $F_c$ of $T^*C$, parameterized by a point $c$ in $C$. These are BAA-type deformations,
as the Hamiltonian for the deformation is the fiber coordinate $p$. 

A second family of BAA branes can be a collection of spectral curves $\Sigma^{(N)}_\mu$ in $T^*C$, a ramified $N$-sheeted cover of $C$ equipped with a flat $U(1)$ bundle, which can be mapped to a line bundle. 
This family is parameterized by the spectral data $\mu$ of a solution of $U(N)$ Hitchin equations on $C$. This is a moduli space of BBB-type deformations, with shape deformations being paired up 
with bundle deformations. The space of ground states on a segment with boundary conditions $F_c$ and $\Sigma^{(N)}_\mu$ clearly gives both the actual $U(N)$ Hitchin system on $C$ 
and the universal bundle for the Hitchin moduli space as we vary $c$ or $\mu$ \cite{witten} 

We can also consider the segment with boundary conditions $\Sigma^{(N)}_\mu$ and $C$ itself. This will give a hyper-holomorphic bundle on the $U(N)$ Hitchin moduli space, 
which we expect to coincide with the Dirac-Higgs bundle in the fundamental representation for $U(N)$. 

Similarly, we can consider a segment with boundary conditions $\Sigma^{(N)}_\mu$ and $\Sigma^{(M)}_{\mu'}$ to get a hyper-holomorphic bundle on the product of $U(N)$ and $U(M)$ 
Hitchin moduli spaces, which we expect to coincide with the Dirac-Higgs bundle in the bi-fundamental representation for $U(N)\times U(M)$.

Similar considerations apply to branes in a complex integrable system $\CM$. For example, the elliptic fibers $F_\mu$ equipped with a flat $U(1)$ bundle 
will have the Langlands dual complex integrable system as a moduli space of BBB-type deformations. Segments with $F_p$ at one end and a generic BAA brane at the other end 
will produce the mirror BBB brane.

Conversely, there should be families of BBB branes on a Hitchin moduli space with a BAA-type deformation space consisting of  bundles for the Langlands dual group,
which implement the opposite Geometric Langlands duality. These should be the BBB branes arising as the BBB-twist of the $T[G]$ theory defined in \cite{Gaiotto:2008ak}. 

\section{The category of BBB branes} \label{app:category}
A BBB brane is a B-brane in each complex structure. A typical mathematical description of B-branes is a derived category of coherent sheaves on the complex manifold. 
We would like to understand better, through general considerations and examples, which general form should such complexes take for a BBB brane. 
Intuitively, we expect to find some sort of complexes of hyper-holomorphic coherent sheaves with $\zeta$-dependent differentials which are holomorphic in every complex structure. 

A simple way to map a B-brane $B$ on a manifold $X$ to an object in the derived category is to consider a segment 
with Dirichlet boundary $D_x$ conditions at one end and $B$ at the other end \footnote{The following argument was developed in the course of an ongoing collaboration with M. Del Zotto and L. Bhardwaj}. This is an $\CN=2$ SQM 
with $x$ as a space of B-type deformations. If we consider a point $x_0$ with a finite gap between the space of 
supersymmetric ground states $V_{x_0}$ and the next excited states, in a sufficiently small neighbourhood of $x_0$ 
we can describe the ground states as the cohomology of a complex $(V_0, \delta_{x_0,x})$ with a differential 
which depends holomorphically on $x$. 

The result is a collection of open sets $U_i$ and complexes of 
holomorphic vector bundles $(V_i, \delta_i)$ defined on $U_i$ equipped with compatible quasi-isomorphisms on intersections of open sets. 
The mathematical term for this structure is a {\it perfect complex}. Perfect complexes are a nice sub-category of the derived category of coherent sheaves.

We can run this analysis equally well for a BBB brane. Now the segment 
with Dirichlet boundary $D_x$ conditions at one end and $B$ at the other end gives us an $\CN=4$ 
SQM with $x$ as a space of BBB-type deformations. If we consider a point $x_0$ with a finite gap between the space of 
supersymmetric ground states $V_{x_0}$ and the next excited states, in a sufficiently small neighbourhood of $x_0$ 
we can describe the ground states by an auxiliary finite-dimensional $\CN=4$ 
SQM with Hilbert space $V_0$ and supercharges $\delta_\alpha$, $\bar \delta_\alpha$. 

The result is some sort of hyper-holomorphic perfect complex: a collection of open sets $U_i$ and $\CN=4$ 
SQM $(V_i, \delta_{\alpha,i})$ with $U_i$ as spaces of BBB-type deformations which have the same supersymmetric ground states on the intersection 
of open sets. It should be possible to improve this description further, possibly with the help of twistorial ideas. 

\section{Gauge group reductions} \label{sec:dir}
We will begin with a discussion of Dirichlet-like boundary conditions and interfaces, which do not require 
adding extra degrees of freedom at the boundary and reduce the gauge group to a subgroup at the boundary. 

This is somewhat orthogonal to the rest of the paper, which focuses on Neumann-like boundary conditions involving 
boundary degrees of freedom coupled to the full gauge group at the boundary. 

Dirichlet-like boundary conditions can actually be recast as Neumann-like boundary conditions
involving specific choices of boundary degrees of freedom. We prefer to review them separately 
as they are already rather well-understood (see e.g. \cite{Witten:2009mh}) and we will need the following results in order to understand some 
important S-duality examples. 

\subsection{Dirichlet boundary conditions}
Half-BPS Dirichlet boundary conditions impose Dirichlet boundary conditions on 
both the connection and the three scalar fields rotated by $SO(3)_C$. The remaining three scalar fields 
rotated by $SO(3)_H$ receive Neumann boundary conditions. Dirichlet boundary conditions admit a 
global symmetry $G_H = G$.

In a BAA twist, the two scalar fields which give rise to the Higgs field on $C$ receive Neumann boundary conditions and are thus 
free to fluctuate at the boundary. The connection on $C$, though, is fixed to some choice of background connection, which can be thought of as a connection for 
the global symmetry group $G_H$. 

In the language of Higgs bundles $(E, \Phi)$, 
the boundary conditions fix the bundle $E$ to some reference $E_H$ and let $\Phi$ unconstrained. Thus the BAA brane is supported on the
complex Lagrangian submanifold in $\CM_H(C,G)$ given by the fiber over a point of the co-tangent bundle to $\Bun(C,G)$.

In the BBB twist, the two scalar fields which give rise to the Higgs field on $C$ receive Dirichlet boundary conditions and are thus 
fixed at the boundary as well. The boundary value can be thought of as a $G_H$ Higgs field. The background connection and Higgs field
must satisfy Hitchin equations on $C$ and give a point $p$ in $\CM_H(C,G)$. The BBB brane is a skyscraper sheaf at $p$. 

Dirichlet boundary conditions can be generalized to interfaces between a $G$ gauge theory and a $G' \subset G$ gauge theory, 
at which the gauge group is reduced from $G$ to $G'$. Such interfaces have a global symmetry group $G_H = Z_{G'}(G)$, the commutant of $G'$ 
in $G$. \footnote{A better characterization would be $N_{G'}(G)/G'$, the quotient of the normalizer of $G'$ in $G$ by $G'$.}

On the BAA side, we restrict the $G$ bundle $E$ to coincide with the $G'$ bundle $E'$ tensored with the reference $G_H$ bundle $E_H$.,
We also require the $G'$ Higgs field $\Phi'$ to coincide with the restriction of the $G$ Higgs field $\Phi$ to $\End(E')$. 
Thus the BAA interface is supported on a simple Lagrangian correspondence. 

On the BBB side, we obtain a sheaf supported on the graph of the embedding of $\CM_H(C,G')$ into $\CM_H(C,G)$
by tensor with a reference solution in $\CM_H(C,G_F)$. We expect the sheaf to be trivial, so that convolution with a sheaf on 
$\CM_H(C,G')$ would give the same sheaf on the image of $\CM_H(C,G')$ in $\CM_H(C,G)$.

\subsection{Nahm pole boundary conditions and interfaces}
Dirichlet boundary conditions admit a surprising modification, which imposes a singular boundary condition on 
the three scalar fields rotated by $SO(3)_H$. The modified boundary condition is labelled by an embedding $\rho$ 
of $\mathfrak{su}(2)$ into the Lie algebra $\mathfrak{g}$ of the gauge group. 
The singular boundary condition breaks the boundary global symmetry to $G_H = Z_\rho(G)$, the commutant of $\rho$ in $G$. 
For a previous discussion of the role of these boundary conditions in Geometric Langlands, see e.g. \cite{Witten:2009mh}.

In the BAA twist, the singular boundary conditions force the Higgs field to lie in a Slodowy slice $S_\rho$. If $t^\pm$ and $t^0$ 
are the images of $\mathfrak{su}(2)$ raising, lowering and Cartan generators respectively, that means 
\begin{equation}
\Phi = t^+ + \Phi^-
\end{equation}
where $[t^-,\Phi^-]=0$. The bundle $E$ is fixed to have the form $E_H \otimes t^0(K_C^{1/2})$, where $E_H$ is some bundle with $G_H$ structure group
and we used $t^0$ to promote $K_C^{1/2}$ to an $SU(2)$ bundle. This is compatible with the restriction on $\Phi$, as $t^\pm$ are global sections of $\End(E)\otimes K_C$.
\footnote{An alternative description of the slice involves a global section $s$ of $\Hom(E_H,E)$ 
with the constraint that $s$, $\Phi s$, $\Phi^2 s$, etc. span flags of a specific form.}

These relations define the support for the 
BAA brane.

If $\rho$ is a maximal embedding, then the support of the BAA brane is simply the canonical section of the Hitchin fibration.
This statement is compatible with S-duality: the S-dual of the maximal Nahm pole boundary condition is a pure Neumann boundary condition, 
whose BBB image is the structure sheaf $\CO$ of the whole Hitchin moduli space. T-duality along the fibers of the Hitchin fibration indeed 
maps $\CO$ to the canonical section of the Hitchin fibration.

The S-dual of more general Nahm pole boundary conditions or pure Dirichlet boundary conditions involves non-trivial boundary degrees of freedom. 

The BBB image of Nahm pole boundary conditions lies somewhat outside the regime of validity of the sigma model description, 
as it involves a singular boundary condition for the scalar fields in the 2d gauge multiplet. It is a variant of the skyscraper sheaf at 
singular loci in the Hitchin moduli space where the Higgs bundle structure group reduces to $G_H$. 

We can promote Nahm pole boundary conditions to interfaces between $G$ and $G' = Z_\rho(G)$ gauge theories. 

On the BAA side that means promoting the Slodowy slice constraints to a Lagrangian correspondence: 
$E_H$ is identified with the bundle $E'$ of the $G'$ Higgs bundle $(E',\Phi')$, while $\Phi'$ is set to equal the projection of $\Phi$ 
onto the trivial $\mathfrak{su}(2)$ representation. 

On the BBB side, that means considering a sheaf supported on the graph of the embedding of $\CM_H(C,G')$ into $\CM_H(C,G)$.
Again, the sigma model description is not quite adequate because of the Nahm pole. 

\bibliography{Langlands}{}
\bibliographystyle{JHEP}
\end{document}